\documentclass[11pt]{article}

\usepackage{amsmath,amsthm,latexsym,amssymb,amsfonts,epsfig}

\numberwithin{equation}{section}
\allowdisplaybreaks

\usepackage{graphicx} 
\usepackage{physics}
\usepackage{amssymb}
\usepackage{color}

\usepackage[hidelinks]{hyperref}
\usepackage{booktabs}

\usepackage{bm}

\allowdisplaybreaks[1] 

\title{{\sf Quantum Field Theory of}\\
{\sf Black Hole Perturbations with Backreaction}\\
{\sf III. Spherically symmetric 2nd order Maxwell sector}}

\author{
{\sf J. Neuser}$^1$\thanks{{\sf jonas.neuser@fau.de}},
{\sf T. Thiemann}$^1$\thanks{{\sf
thomas.thiemann@gravity.fau.de}}\\
\\
{\sf $^1$ Institute for Quantum Gravity, FAU Erlangen -- N\"urnberg,}\\
{\sf Staudtstr. 7, 91058 Erlangen, Germany}\\
}
\date{{\small\sf \today}}

\begin{document}

\maketitle

\begin{abstract}
     In this paper we extend reduced phase space approach to black hole perturbation theory to Maxwell matter. 
     We expand the resulting reduced Hamiltonian to second order in the graviton and photon perturbations  
     and find that the corresponding equations of motion match the ones derived in the literature. Accordingly
     our approach reproduces previous results at second order. Its real virtue lies in the fact that it extends
     to any order in perturbation theory in a manifestly gauge invariant fashion. 
\end{abstract}

\section{Introduction}

In \cite{TT} a reduced phase space approach 
to treat perturbation theory of theories with respect to highly symmetric ``background'' solutions 
of the classical field equations in the presence of gauge redundancies was proposed. 
An application of high interest concerns General Relativity (GR) and its black hole 
solutions where the symmetry group $\cal S$ corresponds to spherical or axi-symmetry and the gauge group 
is the spacetime diffeomorphism group $\cal G$ of which $\cal S$ is a tiny subgroup.

The idea of \cite{TT} is to split the degrees of freedom into four sets as follows: 
The canonical pairs naturally split into two sets 
coined ``symmetric'' and ``non-symmetric'' where the symmetric part is invariant under the action of $\cal S$. 
We then perform a second split of the canonical pairs coined ``gauge'' and ``true'' where the 
symmetric and non-symmetric constraints $(C,Z)$ respectively are solved for chosen 
symmetric and non-symmetric gauge momenta $(p,y)$ respectively. The conjugate symmetric and 
non-symmetric gauge configuration 
variables $(q,x)$ respectively are then subjected to suitable gauge conditions $G$. 
The left over canonical pairs 
then consist of true (i.e. observable) symmetric and non-symmetric canonical pairs 
$(Q,P),(X,Y)$ respectively. Their dynamics is driven by the reduced Hamiltonian $H=H(Q,P,X,Y)$
which 
is obtained as follows: One solves the gauge stability conditions $S:=\{C(f)+Z(g)+B(f,g),G\}=0$ for the 
symmetric and non-symmetric smearing functions $(f,g)$ respectively where $B$ is a 
possible boundary term (if there is a boundary) that ensures that the Poisson bracket is well defined.
Then for any function $F$ on the reduced or ``true'' phase space with coordinates 
$(P,Q,X,Y)$ one requires $S:=\{C(f)+Z(g)+B(f,g),F\}_\ast=0$ where the notation means
that after computing the Poisson bracket we evaluate on the joint solutions $q_\ast, x_\ast$ of 
$G=0$, the solutions $(p^\ast, y^\ast)$ of $C=Z=0$ and the solutions $f_\ast,g_\ast$ of $S=0$.
This ensures that the equations of motion for $F$ driven by $C(f)+Z(g)+B(f,g)$ restricted to 
the reduced phase space are the same as that of $H$.      
 
In this way one can rigorously and non-perturbatively remove all gauge redundancy at the 
classical level. The 
caveat is that for sufficiently complicated systems these steps cannot be carried out explicitly
because one needs to solve PDE's and non-linear algebraic equations. This is 
the point at which perturbation theory comes into play: We consider $X,Y$ as first order 
and $Q,P$ as zeroth order fields and expand the non-perturbative but implicit 
expression $H$ in powers of $X,Y$. Now the function $H$ by construction depends in managable
and explicit form on $Q,P,X,Y,p^\ast,y^\ast$ and it is only due to the intricate dependence 
of $p^\ast,y^\ast$ on $Q,P,X,Y$ that $H$ is not explicitly known. It turns out that 
the expressions $p^\ast,y^\ast$ however can themselves be expanded into powers of $X,Y$
and solved for by an iterative scheme where only ODE's and linear algebraic equations have
to be solved. Therefore $H$ can be explicitly computed pertubatively to any order in the 
true degrees of freedom and no higher order notions of gauge invariance ever have to be 
invented. 

In this way, the reduced system can be treated by the usual methods of 
perturbative Quantum Field Theory where $H$ is truncated at the desired perturbative 
order prior to quantisation. For instance the second order contribution can be used 
to select Fock representations while the higher order terms define the interactions.
This has the attractive feature that interaction or backreaction between 
the true background $Q,P$ and true perturbative degrees of freedom $X,Y$
is faced 
squarely, at least in a perturbative fashion. 

The application of \cite{TT} to black hole physics was outlined in \cite{I}.
The framework is designed to potentially shed new light on the questions of 
backreaction, the mechanism of 
Hawking radiation, black hole evaporation, singularity resolution and 
the corresponding information paradox beyond the usual semiclassical approximation in the sense that 
this is a proper quantum gravity framework where the observable part of the metric, 
encoded among the degrees of freedom $Q,P,X,Y$ is an operator valued distribution 
rather than a classical field.

As argued in \cite{I} a particularly useful gauge condition $G$ is the so called 
Gullstrand-Painlev\'e gauge (GPG) since 1. it corresponds to a foliation of spacetime by 
equal proper time hypersurfaces of free falling observers which therefore come as close 
as possible to define a locally inertial system, 2. the coordinate system 
is regular across the horizon, 3. the intrinsic three metric is 
exactly flat and 4. the spacetime metric is asymptotically flat. In our 
first concrete application of \cite{II} we considered the pure vacuum (i.e. no matter)
sector of the theory for spherical symmetry 
and expanded the reduced Hamiltonian $H$ to second order in $X,Y$.
In that case the degrees of freedom $(Q,P)$ are absent in the sense 
that they reduce to the black hole mass $M$ considered as an integration constant 
while $X,Y$ describe 
tracefree (with respect to the sphere metric) axial (or odd) and polar (or even)
gravitational field polarisations. We found that the Hamiltonian equations of motion 
agree with the second order gauge invariant results found preciously in the literature. 
This confirms the validity of our method whose real virtue lies of course in the ability 
to unambiguously provide the higher order
contributions to $H$.\\
\\
In the present paper we extend \cite{II} by Maxwell matter also in the GPG. Besides 
gravitational waves covered by \cite{II} electromagnetic waves provide another important messenger of 
black hole radiation for astrophysical observation.
For black holes we have access to the region close to the event horizon \cite{Akiyama2019}
and by the usual semi-classical argument the non-vanishing Hawking temperature of the black hole
suggests that black holes 
emit Hawking radiation. Electromagnetic radiation is an important component of the emitted 
spectrum \cite{Page1976a} both by primary and secondary (via particle-antiparticle annihilation)
effects.

The problem of the Maxwell field coupled to gravity has been studied in the literature before. 
For the spherically symmetric background we obtain the charged black hole solution, 
the Reissner Nordstr{\o}m solution. Zerill first studied the  perturbation theory on the level 
of the equations of motion \cite{Zerilli1974}. The Hamiltonian theory was first considered by 
\cite{Moncrief1974a,Moncrief1974RN,Moncrief1975} in Schwarzschild coordinates. A rigorous treatment 
can also be found in the book by Chandrasekhar \cite{Chandrasekhar1983}. 

We demonstrate that we recover the results in the literature from our formalism. 
In contrast to Moncrief we study the Hamiltonian formulation in the Gullstrand-Painlevé coordinates. 
Therefore, there is no divergence in the degrees of freedom across the horizons. 
In the restriction to the spherically symmetric background we recover the Reissner Nordstr{\o}m solution. 
For the perturbations we obtain a physical Hamiltonian describing the dynamics of the gravitational 
and electromagnetic degrees of freedom. The equations of motion for the perturbations match 
the results found in the literature.   

The outline of this manuscript is as follows. In section \ref{sec:GenHamTheory} we present the Hamiltonian 
framework of Einstein -- Maxwell theory. Its exact solutions are derived in section \ref{sec:SymBackground} 
in presence of exact spherical symmetry. In section
\ref{sec:PertSystem} we consider linear perturbations of the exactly symmetric 
background. In section \ref{sec:Reduction} we perform the symplectic reduction and 
derive the reduced Hamiltonian for the observable degrees of freedom. 
Finally, in Section \ref{sec:Comparison} we compare our findings with the ones obtained by Chandrasekhar 
\cite{Chandrasekhar1983} in the Lagrangian formulation. In the appendix we give explicit formulas for some boundary terms that are discussed in the main text.

\section{Hamiltonian Formulation of Einstein -- Maxwell Theory}
\label{sec:GenHamTheory}

The objective of this manuscript is the extension of \cite{II} to include electromagnetic radiation. 
We use the notation of \cite{II} for the gravitational degrees of freedom and implement the
electromagnetic field with the vector potential.

As in the previous paper we work in the Hamiltonian formulation of general relativity. Thus, we study the problem in the ADM approach. We split the metric according to 
\begin{equation}
    \dd{s}^2 = - (N^2 - m_{\mu \nu} N^\mu N^\nu) \dd{t}^2 + 2 m_{\mu \nu} N^\mu \dd{t}\dd{\bm x}^\nu + m_{\mu \nu} \dd{\bm x}^\mu \dd{\bm x}^\nu\,,
\end{equation}
where $N$ is the lapse function, $N^\mu$ is the shift vector and $m_{\mu \nu}$ is the induced metric on the hypersurfaces of the foliation. 

As matter content we consider electromagnetic radiation. We introduce it in the form of a vector potential, i.e. a one-form $A$. For the Hamiltonian formulation we split the vector potential according to $A = A_0 \dd{t} + A_\mu \dd{\bm x}^\mu$. Classically, we describe the dynamics using the Einstein-Hilbert-Maxwell Lagrangian.

The starting point for the model is the full Hamiltonian of the system. It is obtained via a Legendre transformation from the Lagrangian formulation. As in the case for pure gravity this transformation is singular. The Dirac algorithm shows that the quantities $A_0$, $N$ and $N^\mu$ enter the Hamiltonian theory as Lagrange multipliers. Next to the momentum $W^{\mu \nu}$ conjugatre to $m_{\mu \nu}$ we also have the electric field $E^\mu$ which is conjugate to $A_\mu$. 

Compared to the pure gravity case, the Hamiltonian has a similar structure. It is a linear combination of 
modified diffeomorphism and Hamiltonian constraints. Furthermore, we obtain the Gau{\ss} constraint from the 
electromagnetic field. Explicitly, we have
\begin{equation}
    H = \int_\sigma \dd{\Sigma} \qty(N V_0 + N^\mu V_\mu + A_0 V_G)\,,
\end{equation}
where $A_0$, $N$ and $N^\mu$ enter as Lagrange multipliers for the constraints $V_0$, $V_\mu$ and $V_G$. 
It will be sufficient to consider one asymptotic end so that we can focus on the case 
$\sigma=\mathbb{R}_+\times S^2$. There is a new contribution to the Hamiltonian constraint $V_0$:
\begin{equation}
    V_0 = \frac{1}{\sqrt{m}}\qty(m_{\mu \rho} m_{\nu \sigma} - \frac{1}{2} m_{\mu \nu} m_{\rho \sigma})W^{\mu \nu}W^{\rho \sigma}- \sqrt{m} R(m) + \frac{1}{2}\qty(\frac{g^2}{\sqrt{m}} m_{\mu \nu} E^\mu E^\nu + \frac{\sqrt{m}}{2g^2}F_{\mu \nu}F^{\mu \nu})\,,
\end{equation}
where $\sqrt{m}:=\sqrt{\det(m)}$ and $g$ is the coupling constant for the electromagnetic field. We introduced the field strength tensor $F_{\mu \nu}:= \partial_\mu A_\nu - \partial_\nu A_\mu$
and the Ricci scalar $R(m)$ of the three metric $m$. The diffeomorphism constraint also gets a new contribution:
\begin{equation}
    V_\mu = - 2 m_{\mu \rho} D_\nu W^{\nu \rho} + F_{\mu \nu} E^\nu-A_\mu\;V_G
\end{equation}
Finally, the Gau{\ss} constraint $V_G$ is defined as
\begin{equation}
        V_G = \partial_\mu E^\mu
\end{equation}
The conjugate variables $m_{\mu \nu}$ and $W^{\mu \nu}$ as well as $A_\mu$ and $E^\mu$ satisfy the Poisson brackets
\begin{equation}
    \qty{m_{\mu \nu}(t,\bm x),W^{\rho \sigma}(t,\bm y)} = \delta^\rho_{(\mu} \delta^\sigma_{\nu)} \delta(\bm x,\bm y), \quad \quad \quad \qty{A_\nu(t,\bm x), E^\mu(t,\bm y)} = \delta^\mu_\nu \delta(\bm x,\bm y)
\end{equation}
From here on our analysis parallels exactly that of \cite{II}.

\section{Spherical Symmetric Background Solution}
\label{sec:SymBackground}

The parametrization of the spherical symmetric background is unchanged. We use radial 
$x^3=r$ and angular $x^1=\theta,\; x^2=\varphi$ coordinates and denote the metric on 
$S^2$ by $\Omega_{AB},\; A,B=1,2$ with $\det(\Omega)=\sin^2(\theta)$. 
We use
\begin{equation}
    m_{33} = e^{2\mu}, \quad \quad m_{3A} = 0, \quad \quad m_{AB} = e^{2\lambda} \Omega_{AB}\,,
\end{equation}
with two degrees of freedom $\mu$ and $\lambda$. The conjugate momentum is defined as
\begin{equation}
    W^{33} = \sqrt{\Omega} \frac{\pi_\mu}{2}e^{-2\mu}, \quad \quad W^{3A} = 0, \quad \quad W^{AB} = \sqrt{\Omega} \frac{\pi_\lambda}{4}e^{-2\lambda} \Omega^{AB}\,.
\end{equation}
As in the previous paper, we work in the Gullstrand-Painlevé gauge. It is characterized by setting $\mu=0$ and $\lambda=\log(r)$. In this gauge the spherically symmetric degrees of freedom are regular at the horizon. 

The electromagnetic degrees of freedom are also reduced to spherical symmetric ones. 
Since there are no spherically symmetric vector fields on the sphere $S^2$, 
the only non-vanishing component of the vector potential is $A_3$. 
It is an arbitrary function of Gullstrand -- Painlev\'e time $\tau$ and $r$. 
Similarly the only non-vanishing component of the electric field is $E^3$.

Consider first the Gau{\ss} constraint $V_G$. It reduces to $\partial_r E^3 = 0$. 
Therefore, $E^3$ is equal to a constant with respect to the radial coordinate. 
We set $E^3 = \sqrt{\Omega} \xi$. Another simplification comes from the fact that the constraints 
only depend on the vector potential $A_\mu$ through the field strength tensor $F_{\mu \nu}$. 
However, $F_{\mu \nu}$ vanishes in spherical symmetry due to its anti-symmetry. 
In other words, the Hamiltonian is independent of $A_3$. 
This implies through the Hamiltonian equations of motion for $E^3$ that $\xi$ is time-independent. 
As we will see, $\xi$ is related to the electric charge.

Since the field strength tensor is independent of $A_3$ we can freely set it to zero without 
modifying the equations of motion. For reasons of consistency $A_3=0$ needs to be preserved under time 
evolution. This can be achieved by appropriately choosing the Lagrange multiplier field $A_0$.

Inserting the gravitational and electromagnetic degrees of freedom into the constraints we obtain 
a structure for the zeroth order symmetric constraints very similar to the 
vacuum case. Following the notation of \cite{II}, ${}^{(0)}C_v$ gets modified by a new term proportional 
to the square of the electric charge $\xi$:
\begin{align}
    {}^{(0)}C_v &= \frac{4\pi}{8r^2} \qty(\pi_\mu^2 - 2\pi_\mu \pi_\lambda) + 4\pi \frac{g^2}{2r^2}\xi^2\,,\\
    {}^{(0)}C_h &= 4\pi \qty(\frac{1}{r} \pi_\lambda - \pi_\mu')\,.
\end{align}
The plan is to solve these equations for $\pi_\mu$ and $\pi_\lambda$. The method is very similar to the vacuum case. First, we solve ${}^{(0)}C_h$ for $\pi_\lambda$. It is unchanged and reads
\begin{equation}
    \pi_\lambda = r \pi_\mu'\,,
\end{equation}
This can be used in the equation for ${}^{(0)}C_v$ to obtain a differential euqation for $\pi_\mu$:
\begin{equation}
    \pi_\mu^2 - 2 r \pi_\mu \pi_\mu' - 4 r^2 \dv{r}\qty(\frac{g^2 \xi^2}{r} ) = 0,
\end{equation}
For the solution of the equation we multiply by an integrating factor and rewrite it as a total derivative in the following way
\begin{align}
    \dv{r}\qty[\frac{\pi_\mu^2}{r} + 4 \frac{g^2 \xi^2}{r}]=0\,.
\end{align}
In this form the solution is straight forward and we obtain
\begin{align}
    \pi_\mu^2 = 16 rr_s - 4 g^2\xi^2\,,
\end{align}
where $r_s$ is an integration constant which as before is the Schwarzschild radius. As in the vacuum case we can obtain the solution for $\pi_\lambda$ from the Hamiltonian constraint ${}^{(0)} C_v$:
\begin{align}
    \pi_\lambda = \frac{1}{2\pi_\mu}\qty(\pi_\mu^2 + 4 g^2 \xi^2) = \frac{16 r r_s}{2 \pi_\mu}\,.
\end{align}
These formulas provide the zeroth order solutions $\pi_\mu^{(0)},\;\pi_\lambda^{(0)}$ of the symmetric constraints. In the following we will iteratively construct the solution up to second order of the symmetric constraints.

\section{Perturbations of Einstein -- Maxwell Theory}
\label{sec:PertSystem}

We follow the notation of our companion paper \cite{II} and expand the non-symmetric ($l>0$ modes)
into spherical tensor harmonics 
\begin{align}
    m_{33} &= 1 +  \sum_{l\geq 1, m} \bm x^v_{lm} L_{lm}\\
    m_{3A} &= 0 + \sum_{l \geq 1, m,I}\bm x^I_{lm} [L_{I,lm}]_A\\
    m_{AB} &= r^2 \Omega_{AB} + \sum_{l\geq 1,m} \bm x^h_{lm} \Omega_{AB} + \sum_{l \geq 2, m, I} \bm X^I_{lm} [L_{I,lm}]_{AB}\\
    W^{33} &= \sqrt{\Omega}\qty(\frac{\pi_\mu}{2} + \sum_{l\geq 1, m} \bm y^v_{lm} L_{lm})\\
    W^{3A} &= \sqrt{\Omega}\qty(0 +  \frac{1}{2}\sum_{l \geq 1, m,I}\bm y^I_{lm} L^A_{I,lm})\\
    W^{AB} &= \sqrt{\Omega}\qty(\frac{\pi_\lambda}{4r^2}\Omega^{AB} + \frac{1}{2}\sum_{l\geq 1,m} \bm y^h_{lm} \Omega^{AB} + \sum_{l \geq 2, m, I} \bm Y^I_{lm} L^{AB}_{I,lm})
\end{align}
The label $I$ takes the values $e,o$ for the even and odd harmonics respectively. See \cite{II} for the details. We assume the perturbations $\delta W^{\mu \nu}, \delta m_{\mu \nu}$ to follow certain fall-off conditions at infinity. For completeness we recall them here:
\begin{align}
    \label{eq:Decay}
    \begin{split}
    \delta m_{33} &\sim \delta m_{33}^+ r^{-1} + \delta m_{33}^- r^{-2}\\
    \delta m_{3A} &\sim \delta m_{3A}^+ + \delta m_{3A}^- r^{-1}\\
    \delta m_{AB} &\sim \delta m_{AB}^+ r + \delta m_{AB}^-\\
    \delta W^{33} &\sim \delta W^{33}_- + \delta W^{33}_+ r^{-1}\\
    \delta W^{3A} &\sim \delta W^{3A}_- r^{-1} + \delta W^{3A}_+ r^{-2}\\
    \delta W^{AB} &\sim \delta W^{AB}_- r^{-2} + \delta W^{AB}_+ r^{-3}\,.
    \end{split}
\end{align}
The quantities $\delta W^{\mu \nu}_\pm,\delta m_{\mu\nu}^\pm$ on the right hand side are independent of the radial coordinate $r$ but still have angular dependence through $l,m$. The sub-/superscript $\pm$ stands for the behaviour of the variables under parity transformations. Details on this and the relation to the notion of parity in the rest of the manuscript can be found in \cite{II}.

Similarly to the gravitational degrees of freedom we expand the perturbed vector potential $A_\mu$ and the perturbed electric field $E^\mu$ into tensor harmonics . We use the convention
\begin{align}
    A_3 &= \sum_{l\geq 1, m} \bm x_M^{lm} L_{lm}\\
    A_B &= \sum_{l\geq 1, m,I} \bm X_M^{I,lm} [L_{I,lm}]_B\\
    E^3 &= \sqrt{\Omega}  \qty(\xi + \sum_{l\geq 1, m}\; \bm y^M_{lm} L_{lm})\\
    E^B &= \sqrt{\Omega} \sum_{l\geq 1, m,I}\
    \bm Y^M_{I,lm} L_{I,lm}^B 
\end{align}
The sub- or superscript $M$ denotes ``Maxwell'' degrees of freedom and $x,y$ as before denote 
non-symmetric gauge degrees of freedom while $X,Y$ denote non-symmetric true degrees of freedom.

We choose the fall-off conditions of the electromagnetic perturbations as
\begin{align}
    \label{eq:FAllOffEM}
    \begin{split}
    \delta A_3 &\sim \delta A_3^+ r^{-1} + \delta A_3^- r^{-2}\\
    \delta A_B &\sim \delta A_B^+ + \delta A_B^- r^{-1}\\
    \delta E^3 &\sim \delta E^3_- + \delta E^3_+ r^{-1}\\
    \delta E^B &\sim \delta E^B_- r^{-1} + \delta E_B^+ r^{-2}
    \end{split}
\end{align}
As in the gravitational case the variables $\delta A_\mu^\pm$ and $\delta E^\mu_\pm$ are constants with respect to $r$ but are still allowed to vary with $l,m$. The sub-/superscript $\pm$ stands again for the parity of these perturbations.

In the next step we expand the Hamiltonian, diffeomorphism and Gau{\ss} constraints to second order and extract the symmetric and non-symmetric contributions.

\subsection{Perturbations and exact solutions of the Gau{\ss} constraint}

Before studying the diffeomorphism and the 
Hamiltonian constraints we consider the Gau{\ss} constraint. Similarly to the spherically symmetric background 
it simplifies the analysis to first solve this constraint. It only has a first order correction which is given by
\begin{equation}
    ({}^{(1)}V_G)_{lm} = \sqrt{\Omega}((\bm y^M_{lm})' - \sqrt{l(l+1)}\bm Y^M_{e,lm})
\end{equation}
The gauge degrees of freedom are $\bm x_M, \bm y^M$ while the true degrees of freedom are $X_M, Y^M$. Hence, we solve the Gau{\ss} constraint for $\bm y^M$ and have
\begin{equation}
    \bm y^M_{lm} = \sqrt{l(l+1)} \int \bm Y^M_{e,lm}\dd{r}\,.
\end{equation}
We fix the gauge of the electromagnetic field by setting $\bm x_M^{lm}=0$. Since there are no higher order corrections to the Gau{\ss} constraint we have solved $V_G=0$ to all orders.

\subsection{Second order perturbations of spatial diffeomorphism and Hamiltonian constraints}

As in the vacuum case we 
only need the first oder, automatically non-symmetric, and second order symmetric constraints 
in order to compute the reduced Hamiltonian to second order.\\
\\
The first order non-symmetric constraints are
\begin{align}
    {}^{(1)}Z^h_{lm} &= -2 \partial_r \bm y_v + \sqrt{l(l+1)} \bm y_e + 2 r \bm y_h - \partial_r \pi_\mu \bm x^v - \frac{1}{2} \pi_\mu \partial_r \bm x^v +\frac{\pi_\lambda}{2 r^2} \sqrt{l(l+1)} \bm x^e + \frac{\pi_\lambda}{2 r^2} \partial_r \bm x^h\\
    {}^{(1)}Z^e_{lm} &=\sqrt{2(l+2)(l-1)}\qty(r^2 \bm Y_e+ \frac{\pi_\lambda}{4r^2} \bm X^e) -\partial_r(r^2 \bm y_e + \pi_\mu \bm x^e) + \! \sqrt{l(l+1)} \qty(\frac{\pi_\mu}{2}  \bm x^v -r^2 \bm y_h) -\! \xi \partial_r \bm X^e_M \\
    {}^{(1)}Z^o_{lm} &= \sqrt{2(l+2)(l-1)}\qty(r^2 \bm Y_o + \frac{\pi_\lambda}{4 r^2}\bm X^o) - \partial_r \qty(r^2 \bm y_o + \pi_\mu \bm x^o) - \xi \partial_r \bm X^o_M\\
    \begin{split}
    {}^{(1)}Z^v_{lm} &=  \frac{1}{2r^2} (\pi_\mu - \pi_\lambda) \bm y_v  - \qty(2 r \partial_r + l(l+1) + 2 - 2 \frac{r_s}{r}) \bm x^v- \frac{1}{2} \pi_\mu \bm y_h-\frac{1}{r^2}\sqrt{\frac{(l+2)(l+1)l(l-1)}{2}} \bm X^e\\
    &+ 2\qty( \partial_r^2 - \frac{1}{r} \partial_r - \frac{(l+2)(l-1)}{2r^2} - \frac{r_s}{r^3}) \bm x^h + 2  \sqrt{l(l+1)} \qty(\partial_r + \frac{1}{r}) \bm x^e + \frac{g^2}{r^2} \xi \sqrt{l(l+1)} \int \bm Y^M_{lm} \dd{r}
    \end{split}
\end{align}

The second order symmetric constraints are given by 
\begin{align} \label{2nd}
    \begin{split}
    {}^{(2)}C_h &= - \bm x^o \cdot \partial_r \bm y_o + \bm Y_o \cdot \partial_r \bm X^o + \bm y_v \cdot \partial_r \bm x^v - 2 \partial_r (\bm x^v \cdot \bm y_v) - \bm x^e \cdot \partial_r \bm y_e + \bm Y_e \cdot \partial_r \bm X_e + \bm y_h \cdot \partial_r \bm x^h\\
    &~~~+ \bm Y^M_{e} \cdot (\bm X^e_M)' + \bm Y^M_o \cdot (\bm X^o_M)'
    \end{split}\\
    {}^{(2)}C_v &= \frac{1}{2} \bm y_o \cdot \bm y_o + \frac{1}{2r^2}\pi_\mu \bm x^o \cdot \bm y_o + \frac{1}{r^2}\bm x^o \cdot \qty(4r \partial_r - 3 + \frac{l(l+1)}{2} + 2 \frac{r_s}{r})\bm x^o + r^2 \bm Y_o \cdot \bm Y_o\nonumber\\
    &- \frac{1}{2r^2}(\pi_\mu - \pi_\lambda) \bm Y_o \cdot \bm X^o - \frac{1}{r^4}\bm X^o\cdot \qty(r^2 \partial_r^2 - 4 r \partial_r  + \frac{7}{2} + \frac{r_s^2}{g^2 \xi^2 - 4 r r_s})\bm X^o - \frac{3}{4r^2} \partial_r \bm X^o \cdot \partial_r \bm X^o\nonumber\\
    &-\frac{1}{r^3} \sqrt{\frac{(l+2)(l-1)}{2}}\bm x^o\cdot \qty( r\partial_r -2) \bm X^o\nonumber\\
    &+\frac{1}{2 r^2} \bm y_v \cdot \bm y_v + \frac{1}{4 r^2}(3 \pi_\mu  - \pi_\lambda)\bm x^v \cdot \bm y_v + \bm x^v \cdot \qty(3 r \partial_r + 1 + \frac{r_s}{r} - \frac{g^2 \xi^2}{4r^2})\bm x^v + \frac{1}{2} \bm y_e \cdot \bm y_e + \frac{1}{2 r^2} \pi_\mu \bm x^e \cdot \bm y_e\nonumber\\
    &+ \frac{1}{r^2}\bm x^e \cdot \qty(4 r \partial_r - 3 + 2\frac{r_s}{r})\bm x^e - \frac{1}{2r^4}\qty(4\qty(1 -  \frac{r_s}{r})\bm x^h \cdot \bm x^h  - 4 r \bm x^h \cdot \partial_r \bm x^h + r^2 \partial_r \bm x^h \cdot  \partial_r \bm x^h)\nonumber\\
    &+ r^2 \bm Y_e \cdot \bm Y_e - \frac{1}{2 r^2}  (\pi_\mu - \pi_\lambda) \bm X^e \cdot \bm Y_e - \frac{1}{r^4} \bm X^e \cdot \qty(r^2 \partial_r^2 - 4 r \partial_r + \frac{7}{2} + \frac{r_s^2}{g^2 \xi^2 - 4 r r_s})\bm X^e\\
    &- \frac{3}{4 r^2} \partial_r \bm X^e \cdot \partial_r \bm X^e - \bm y_h \cdot \bm y_v - \frac{\pi_\mu}{2 r^4}\bm y_v \cdot \bm x^h - \frac{1}{4}\pi_\mu \bm x^v \cdot \bm y_h - \partial_r \bm x^h \cdot \partial_r \bm x^v\nonumber \\
    &- \frac{1}{r^2} \bm x^v \cdot \qty( r^2 \partial_r^2 \bm x^h - r \partial_r \bm x^h + \frac{l(l+1)+2}{2} \bm x^h + 3 \frac{r_s}{r} \bm x^h - \frac{g^2 \xi^2}{2 r^2} \bm x^h) - \sqrt{l(l+1)}\bm x^v \cdot \partial_r \bm x^e\nonumber\\
    &- \frac{1}{2 r^2}\sqrt{\frac{(l+2)(l+1)l(l-1)}{2}} \bm x^v \cdot \bm X^e -  \sqrt{l(l+1)} \frac{1}{r}\bm x^e \cdot \qty( - \bm x^v + r \partial_r \bm x^v)\nonumber\\
    &+ \frac{1}{r^3}\sqrt{l(l+1)} \bm x^e \cdot \qty(2 \bm x^h - r \partial_r \bm x^h) + \sqrt{\frac{(l+2)(l-1)}{2}} \frac{1}{r^3}\bm x^e \cdot \qty( 2 \bm X^e - r \partial_r \bm X^e)\nonumber\\
    & + \frac{g^2}{2r^2}\Bigg[\qty(2 \bm x^o \cdot \bm Y^M_o + \sqrt{l(l+1)} \qty(\bm x^v  - \frac{2}{r^2} \bm x^h ) \cdot \int \bm Y^M_{e}\dd{\tilde r} + 2 \bm x^e \cdot \bm Y^M_{e})\xi\\
    &~~~~~~~~~~~~+ l(l+1)\int \bm Y^M_{e}\dd{\tilde r}\cdot \int \bm Y^M_{e}\dd{\tilde r} + r^2 (\bm Y^M_{e} \cdot \bm Y^M_e+\bm Y^M_o \cdot \bm Y^M_o) \Bigg]\nonumber\\
    &+ \frac{1}{2g^2 r^2} \qty[l(l+1)\bm X^o_M \cdot \bm X^o_M + r^2 (\bm X^o_M{}' \cdot \bm X^o_M{}' + \bm X_M^e{}'\cdot \bm X^e_M{}')]\nonumber
\end{align}

\section{Reduced phase space formulation}
\label{sec:Reduction}

We are now fully prepared to analyse the constraints and to obtain the physical Hamiltonian. 
The procedure is similar to the vacuum case with some changes of the transformations due to the electromagnetic field. 
As before we start with the analysis of the second order constraints. 
After that, we use the first order constraints to obtain the reduced Hamiltonian. For the odd parity first order constraints we follow the programme and solve the constraint $Z^o$ for the gauge degrees of freedom $\bm x^o,\bm y_o$. 
For the even parity we proceed differently. In an intermediate step we solve the constraints for $\bm x^h$, $\bm y^e$ and $\bm Y^e$ in the gauge $\bm x^e=0$, $\bm X^e=0$ and $p_2=0$, where $p_2$ is a new variable that we will introduce.
The reason for this is that for this setup the calculation is easier. 
In the end of the section we shopw that this is equivalent to solving the first order constraints in the other gauge ($\bm x^v=\bm x^h= \bm x^e=0$) and keeping $\bm X^e$ and $\bm Y_e$ as true degrees of freedom. 
As in our companion paper \cite{II} the final description drastically simplifies after 
performing suitable 
canonical transformations.

\subsection{Solution of the second order constraints}

We solve the second order Hamiltonian and diffeomorphism constraints exactly the same way as 
in the vacuum case. We assume that the first order constraints are solved and we split the 
symmetric momenta into the background and the contributions second order in the perturbations: $\pi_\mu^{(0)} + \pi_\mu^{(2)}$ and $\pi_\lambda^{(0)} + \pi_\lambda^{(2)}$. Then, the symmetric constraints to second order are
\begin{align}
    C_v &\sim \frac{4\pi}{4 r^2} \qty(\pi_\mu^{(0)}\pi_\mu^{(2)} - \pi_\mu^{(0)} \pi_\lambda^{(2)} - \pi_\mu^{(2)} \pi_\lambda^{(0)}) + {}^{(2)}C_v=0\,,\\
    C_h &\sim 4\pi \qty(\frac{1}{r} \pi_\lambda^{(2)} - (\pi_\mu^{(2)})' + {}^{(2)}C_h)=0\,.
\end{align}

The terms $ {}^{(2)}C_v$ and ${}^{(2)}C_h$ are the contributions to the constraints which are of second order 
in the perturbations $X,Y$. They will be studied in detail later after we solved the first order constraints
and are given by inserting into (\ref{2nd}) the first order solutions $y^\alpha(1)$ to the first order constraint 
equations for $y^\alpha,\; \alpha=v,h,e,o$
in terms of $X,Y$. They do not depend on $\pi_\mu^{(2)},\pi_\lambda^{(2)}$ and 
$\pi_\mu^{(1)}=\pi_\lambda^{(1)}=y^\alpha(0)=0$ by construction. 
 
The structure of the second order constraints is precisely the same as in \cite{II}. 
Therefore, we can use the results obtained there and have
\begin{align}
    \pi_\mu^{(2)} = \frac{4 r}{4 \pi \pi_\mu^{(0)}} \int \dd{r} \qty[\frac{\pi_\mu^{(0)} }{4 r} {}^{(2)} C_h + {}^{(2)}C_v]
    \label{eq:SolIntegralpimu2}
\end{align}
We use the equation $C_v=0$ to get the solution for $\pi_\lambda^{(2)}$. It is given by
\begin{equation}
    \pi_\lambda^{(2)} = \qty(1 - \frac{\pi_\lambda^{(0)}}{\pi_\mu^{(0)}}) \pi_\mu^{(2)} + \frac{4 r^2}{4 \pi \pi_\mu^{(0)}}{}^{(2)}C_v
\end{equation}
For the physical Hamiltonian we need the second order expansion for $\pi_\mu$. We have
\begin{align}
    \pi_\mu \sim \pi_\mu^{(0)} + \pi_\mu^{(2)} =\pi_\mu^{(0)}\qty[ 1 + \frac{1}{4\pi(4 r_s - g^2 \xi^2 / r)}
    \int \dd{r}\qty(\sqrt{\frac{r_s}{r} - \frac{g^2 \xi^2}{4 r^2}} {}^{(2)}C_h + {}^{(2)}C_v)]
\end{align}

\subsection{The dipole perturbations \texorpdfstring{$(l=1)$}{(l=1)}}

We treat the dipole perturbations separately because the spherical tensor harmonics are not defined for $l=1$. In the previous paper \cite{II} we saw that the constraints can be solved explicitly for the gravitational degrees of freedom. There was no dynamics for the dipole perturbations. In the computation appeared integration constants that we related to viewing the Schwarzschild metric in an accelerated frame of reference. 

In the presence of matter such as the electromagnetic field we have true degrees of freedom already for $l=1$. In fact for our model these degrees of freedom are $\bm X^{e/o}_M, \bm Y_{e/o}^M$. Later we expect the physical Hamiltonian to depend on these degrees of freedom and to describe the dynamics of the $l=1$ modes of the  electromagnetic field. Additionally, we anticipate to see some integration constants describing the charged black hole solution in an accelerated frame of reference.

In the following we solve the constraints for the even and odd parity sector separately and derive the solution for $\pi_\mu^{(2)}$. For the odd parity sector we have one first order constraint, $Z^o$. The solution of the differential equation in the gauge $\bm x^o=0$ is given by
\begin{equation}
    \bm y_o^{1m} = \frac{1}{r^2}\qty(a_m + \xi \bm X^{o,1m}_M)\,.
\end{equation}
Exactly as in the gravitational case we introduced an integration constant $a_m$. The solution $\bm y_o^{1m}$ is now plugged into the formula \eqref{eq:SolIntegralpimu2} for $\pi_\mu^{(2)}$. The odd parity contribution for $l=1$ is
\begin{align}
    \frac{\pi \pi_\mu^{(0)}}{r}\pi_\mu^{(2)}\Big|_{l=1,odd} = \sum_{m} \int \dd{r} \Big[&N^3 \bm Y^M_{o,1m} \bm X^M_{o,1m}{}' + N\Big(\frac{g^2}{2} (\bm Y^M_{o,1m})^2 + \frac{1}{2g^2}\qty((\bm X^{o,1m}_M{}')^2 + \frac{2 + g^2 \xi^2 r^{-2}}{r^2} (\bm X^{o,1m}_M)^2)\nonumber\\
    &+\frac{a_m^2 + 2 \xi a_m \bm X^{o,1m}_M}{2 r^4} \Big)\Big]
\end{align}
We will discuss the physics of the odd $l=1$ modes of the electromagnetic field later when we have access to the physical Hamiltonian.

For the even parity perturbations we have three first order constraints $Z^v, Z^h, Z^e$ that we solve for $\bm y_h, \bm y_v, \bm y_e$. The true degrees of freedom are the modes $\bm X^{e,1m}_M$ and $\bm Y_{e,1m}^M$ of the electromagnetic field. We start with the solution of $Z^v=0$ for $\bm y_h$. This gives
\begin{equation}
    \bm y_h^{1m} = \frac{1}{r^2}\qty(1 - \frac{\pi_\lambda}{\pi_\mu}) \bm y_v^{1m} + \frac{2 \sqrt{2} g^2 \xi}{r^2 \pi_\mu} \int \bm Y_{e,1m}^M \dd{r}\,.
\end{equation}
Then, the solution of $Z^h=0$ for $\bm y_e$ gives
\begin{equation}
    \bm y_e^{1m} = \frac{1}{\sqrt{2}} \qty(2 \partial_r \bm y^{1m}_v - \frac{2}{r}\qty(1 - \frac{\pi_\lambda}{\pi_\mu}) \bm y^{1m}_v) - \frac{4 g^2 \xi}{r \pi_\mu} \int \bm Y_{e,1m}^M \dd{r}
\end{equation}
The last constraint $Z^e$ together with the solutions for $\bm y_h$ and $\bm y_e$ leads to a differential equation for $\bm y_v$:
\begin{equation}
    \sqrt{2} r^2 \partial_r^2 \bm y_v^{1m} + \frac{\sqrt{2} r(6 r r_s - g^2 \xi^2)}{4 r r_s - g^2 \xi^2} \partial_r \bm y^{1m}_v - \frac{2 \sqrt{2} g^2 \xi^2 r r_s}{(4 r r_s - g^2 \xi^2)^2} \bm y^{1m}_v =s(r)\,,
\end{equation}
with a ``source'' term $s(r)$ depending on the electromagnetic field:
\begin{align}
    s(r)&=\xi \partial_r \bm X^{e,1m}_M - \frac{4 g^2 \xi r r_s}{(4 r r_s - g^2 \xi^2)^{3/2}}\int \bm Y^M_{e,1m}{}\dd{r} + \frac{2 g^2 \xi r}{\sqrt{4 r r_s- g^2 \xi^2}} \bm Y^M_{e,1m}\,.
\end{align}
For an uncharged black hole ($\xi=0$) the source term vanishes.
In the discussion of the solution of the differential equation in the following paragraphs we are not displaying the labels $l=1,m$.

The differential equation for $\bm y_v$ is an inhomogeneous second order linear differential equation. The general solution of this equation is given by a general solution of the homogeneous equation and a particular solution of the inhomogeneous equation. The homogeneous solution is a linear combination $\bm y_v = C_I\bm y_v^I + C_{II}\bm y_v^{II}$ of two independent solutions $\bm y_v^{I}, \bm y_v^{II}$ given by 
\begin{align}
    \bm y_v^I &= \frac{1}{\pi_\mu}\\
    \bm y_v^{II} &= 1- \frac{2 g \xi}{\pi_\mu}\arctan(\frac{\pi_\mu}{2 g \xi})
\end{align}

We use the information about the homogeneous solution to derive a particular solution of the inhomogeneous equation. We will use the method of variation of constants. In the computation we consider the constants $C_I$ and $C_{II}$ to be functions of $r$, i.e.
\begin{align}
    \bm y_v^{\mathrm{part}}(r) = C_I(r) \bm y_v^{I}(r)+ C_{II}(r) \bm y_v^{II}(r)
\end{align}
We insert this ansatz into the differential equation including the source term $s(r)$. Then, we have the equation 
\begin{align}
    \sqrt{2} r^2 (C_{I}'' \bm y_v^{I} + 2 C_{I}' \bm y_v^{I}{}' +C_{II}'' \bm y_v^{II} + 2 C_{II}' \bm y_v^{II}{}') + \frac{\sqrt{2} r(6 r r_s - g^2 \xi^2)}{4 r r_s - g^2 \xi^2} (C_{I}' \bm y_v^{I}+C_{II}' \bm y_v^{II}) = s(r)\,.
\end{align}
The terms with no derivatives of $C_I$ and $C_{II}$ vanish because $\bm y_v^I$ and $\bm y_v^{II}$ are solutions to the homogeneous equation. The above equation is satisfied if $C_I$ and $C_{II}$ satisfy the differential equations
\begin{align}
    C_{I}' \bm y_v^{I}+C_{II}' \bm y_v^{II} &=0\,,\\
     \sqrt{2} r^2 \qty(C_{I}' \bm y_v^{I}{}' + C_{II}' \bm y_v^{II}{}') &= s(r)\,.
\end{align}
The first is used to replace $\bm y_v^{II}$ in the second equation 
\begin{align}
    \sqrt{2} r^2 C_I{}' \qty(\bm y_v^{I}{}' -  \frac{\bm y_v^I}{\bm y_v^{II}} \bm y_v^{II}{}') &= s(r)
\end{align}
Therefore
\begin{align}
    C_I = \int \frac{s(r)}{\sqrt{2} r^2\qty(\bm y_v^{I}{}' -  \frac{\bm y_v^I}{\bm y_v^{II}} \bm y_v^{II}{}')} \dd{r}\,.
\end{align}
Similarly we obtain 
\begin{align}
    C_{II} = \int \frac{s(r)}{\sqrt{2} r^2\qty(\bm y_v^{II}{}' -  \frac{\bm y_v^{II}}{\bm y_v^{I}} \bm y_v^{I}{}')} \dd{r}\,.
\end{align}
Then the particular solution is 
\begin{align}
    \bm y_v^{\mathrm{part}} =  \int \frac{s(r)}{\sqrt{2} r^2\qty(\bm y_v^{I}{}' -  \frac{\bm y_v^I}{\bm y_v^{II}} \bm y_v^{II}{}')} \dd{r} \bm y_v^{I}+ \int \frac{s(r)}{\sqrt{2} r^2\qty(\bm y_v^{II}{}' -  \frac{\bm y_v^{II}}{\bm y_v^{I}} \bm y_v^{I}{}')} \dd{r} \bm y_v^{II}\,.
\end{align}
The solution is now inserted into the equation for $\pi_\mu^{(2)}$. We obtain
\begin{align}
    \frac{\pi \pi_\mu^{(0)}}{r}\pi_\mu^{(2)}\Big|_{l=1,\text{even}}&=\sum_m\int \dd{r} \frac{\pi_\mu}{4r} \bm Y_{e,1m}^M (\bm X^{e,1m}_M)' + \frac{1}{2} (\bm y^{1m}_e)^2 - \qty(\frac{2 \sqrt{2} g^2 \xi}{r^2 \pi_\mu} \int \bm Y^M_{e,1m}\dd{r}) \bm y_v + \frac{1}{2g^2}\qty[\bm X^{e,1m}_M{}']^2\nonumber\\
    &+ \frac{g^2}{2r^2}\qty(2 \qty(\int \bm Y_{e,1m}^M \dd{r})^2 + r^2 (\bm Y_{e,1m}^M)^2)\,,
\end{align}
where we need to replace $\bm y_e$ and $\bm y_v$ by the corresponding solutions of the differential equation. It is convenient to remove the integral of $\bm Y^e_M$ and the derivative of $\bm X^e_M$ with the following canonical transformation where we introduce new variables $A^e$ and $\Pi^e_A$ defined as
\begin{align}
    A^{e,1m} &= g^2 \int \bm Y_{e,1m}^M \dd{r}\,,\\
    \Pi_A^{e,1m} &= g^{-2}\partial_r \bm X^{e,1m}_M\,.
\end{align}
In the new variables we have
\begin{align}
    \begin{split}
    \frac{\pi \pi_\mu^{(0)}}{r}\pi_\mu^{(2)}\Big|_{l=1,\text{even}}=\sum_m\int \dd{r} &\frac{\pi_\mu}{4r}\Pi_A^{e,1m} A^{e,1m}{}' + \frac{1}{2} (\bm y^{1m}_e)^2 - \frac{2 \sqrt{2} \xi}{r^2 \pi_\mu}A^{e,1m} \bm y^{1m}_v + \frac{g^2}{2}(\Pi^{e,1m}_A)^2\\
    &+ \frac{1}{2g^2r^2}\qty(2 (A^{e,1m})^2 + r^2 (A^{e,1m}{}')^2) \,.
    \end{split}
\end{align}

\subsection{Solution of the first order constraints - odd parity}

In the odd parity sector there is one first order constraint which we are now using to eliminate the gauge degrees of freedom $\bm x^o$ and $\bm y_o$. We are left with two pairs of true degrees of freedom. In the electromagnetic sector we have $\bm X^o_M, Y_o^M$ and in the gravitational sector we have $\bm X^o, \bm Y_o$. 

The solution of the constraint equation $Z^o=0$ for $\bm y_o$ is 
\begin{align}
    \bm y_o^{(1)} = \frac{1}{r^2}\int \qty[\sqrt{2(l+2)(l-1)}\qty(r^2 \bm Y_o + \frac{\pi_\lambda^{(0)}}{4 r^2}\bm X^o) - \xi \partial_r \bm X^o_M]\dd{r}\,.
\end{align}
This is all that needs to be done at first order. The solution for $\bm y_o$ is now inserted into the formula for $\pi_\mu^{(2)}$ in equation \eqref{eq:SolIntegralpimu2}. However, the resulting function is still lengthy but can be simplified using canonical transformations. This is the goal of the rest of this subsection. 

For the electromagnetic sector the canonical transformation is trivial. However, to unify the notation with the treatment in the even parity sector we redefine $A^o = \bm X^o_M$ and $\Pi_A^o = \bm Y_o^M$. In the gravitational sector we define new quantities $Q,P$ which are defined as
\begin{align}
    P &:= \frac{1}{\sqrt{2}} \partial_r \qty(r^{-2} \bm X^o)\\
    Q &:= \sqrt{2} \int \dd{r}\qty( r^2 \bm Y_o + \frac{\pi_\lambda^{(0)}}{4 r^2} \bm X^o)\,.
\end{align}
This transformation has the same structure as in the case without any electromagnetic field. The only difference is that now $\pi_\lambda$ depends on the background electric charge $\xi$.

Everything we have developed so far in this section is now inserted into equation \eqref{eq:SolIntegralpimu2}. That is we replace $\bm y_o$ by its solution $\bm y_o^{(1)}$ and fix the gauge by setting $\bm x^o=0$. We also substitute the variables $\bm X^o,\bm Y_o$ and $\bm X^o_M, \bm Y_o^M$ by the new quantities $Q,P$ and $A_o, \Pi_A^o$ respectively. The resulting expression is further simplified using integration by parts. We end up with 
\begin{align}
\begin{split}
    \frac{4r}{\pi_\mu^{(0)}} \pi_\mu^{(2)}\Big|_{l\geq 2,\text{odd}} = \int \dd{r} &\frac{1}{4 r}\pi_{\mu}^{(0)}(P Q' + \Pi_A^o A^o{}') + \frac{1}{2}\qty(r^2 P^2 + \frac{1}{r^4}(l+2)(l-1) Q^2 + \frac{1}{r^2} (Q')^2) \\
    &+ \frac{1}{2} \qty(g^2 (\Pi_A^o)^2 + \frac{1}{g^2}\qty(\frac{l(l+1)}{r^2} + \frac{g^2 \xi^2}{r^4}+({A^o}')^2 )) -  \frac{\sqrt{(l+2)(l-1)}}{r^4} \xi Q A^o
\end{split}
\end{align}
The boundary term that we dropped in the computation is given by
\begin{align}
    \int \dd{r} \dv{r} \qty(2 r^2 P \int P \dd{r} + \frac{1}{2}(2r +r_s) \qty(\int P \dd{r})^2)\,.
\end{align}
For the physical Hamiltonian that we introduce later we are interested in the limit $r$ to infinity of $\pi_\mu^{(2)}$. We will now show that in this limit the boundary term is not contributing. From the fall-off behaviour of the canonical fields in equation \eqref{eq:FAllOffEM} we deduce that $Q$ grows at most linearly in $r$ and $P$ decays as $r^{-2}$. Then, the boundary term behaves as $r^{-1}$. This means that it vanishes in the limit and the boundary term can be dropped.

We implement another canonical transformation to change the factors of $r$ in $\pi_\mu^{(2)}$ and to introduce the equivalent of the Regge-Wheeler potential into the Hamiltonian. For this we introduce $Q^o$ and $P^o$ defined by a rescaling of $Q$ and $P$ by $r$ and a shift of $P$. We have
\begin{align}
    Q &= r Q^o\\
    P &= \frac{1}{r} \qty(P^o - \frac{\pi_\mu}{4r^2} Q^o)
\end{align}
Inserting this into the solution for $\pi_\mu^{(2)}$ we obtain

\begin{align}
\begin{split}
    \frac{4r}{\pi_\mu^{(0)}} \pi_\mu^{(2)}\Big|_{l\geq 2,\text{odd}} =& \int \dd{r}\Bigg[\frac{1}{4 r}\pi_{\mu}^{(0)}(P^o (Q^{o})' + \Pi_A^o (A^o)') -  \frac{\sqrt{(l+2)(l-1)}}{r^3} \xi Q^o A^o\\
    &+ \frac{1}{2} \Big((P^o)^2 + (Q^o{}')^2 + \frac{1}{r^4}(l(l+1)r^2 - 3 r r_s + g^2 \xi^2)(Q^{o})^2\Big)\\
    &+ \frac{1}{2}\qty( g^2(\Pi_A^o)^2 + \frac{1}{g^2}\qty((A^o{}')^2 + \frac{1}{r^4}\qty(l(l+1)r^2 + g^2\xi^2 )(A^o)^2))\Bigg]\,.
\end{split}
\end{align}
In the calculation we used integration by parts to simplify the resulting expression. The boundary term that we dropped is given by
\begin{align}
    - \int \dd{r} \dv{r}\qty(\frac{1}{2 r^2}\qty(r_s - r - \frac{g^2 \xi^2}{4r}) (Q^o)^2)\,.
\end{align}
We showed earlier that $Q$ grows at most linearly in $r$. Thus, $Q^o$ will 
approach a constant as $r$ tends to infinity. Then, the boundary 
term vanishes as $r^{-1}$ in the limit $r \to \infty$.

In view of the results for the even parity degrees of freedom, we introduce the gravitational potential $V^o_{\mathrm{grav}}$, the electromagnetic potential $V^o_{\mathrm{em}}$ and the coupling potential $V^o_{\mathrm{Coup}}$.
\begin{align}
    V_{\mathrm{grav}}^{o} &=\frac{1}{r^2}\qty(U^o - \frac{3 r_s}{2 r} W^o)\\
    V_{\mathrm{em}}^{o} &=\frac{1}{r^2}\qty(U^o + \frac{3 r_s}{2 r}W^o)\\
    V_{\mathrm{Coup}}^{o} &=\frac{g \xi}{r^3} W^o\,.
\end{align}
The potentials depend on the quantities $W^o$ and $U^o$ defined as
\begin{align}
    W^{o} &= -1\\
    U^{o} &= l(l+1) + \frac{3 r_s}{2 r} + \frac{g^2 \xi^2}{r^2}\,.
\end{align}
We use the potentials in the integral for the odd parity variables. Then, we have the first intermediate result for the odd parity sector:
\begin{align}
    \frac{4r}{\pi_\mu^{(0)}} &\pi_\mu^{(2)}\Big|_{l\geq 2,\text{odd}}=\int \dd{r} \frac{1}{4r}\pi_{\mu}^{(0)}(P^o (Q^{o})' + \Pi_A^o (A^o)')\\
    &+ \frac{1}{2} \qty((P^o)^2 + g^2(\Pi_A^o)^2 + (Q^o{}')^2 +\frac{1}{g^2}(A^o{}')^2 + V_{\mathrm{grav}}^{o} (Q^{o})^2 + \frac{1}{g^2} V_{\mathrm{em}}^{o} (A^o)^2 + \frac{2}{g} \sqrt{(l+2)(l-1)} V_{\mathrm{Coup}}^{o} Q^o A^o)\nonumber\,,
\end{align}

\subsection{Solution of the first order constraints - even parity}

The even parity sector is significantly more complicated compared to the odd parity. We have to solve three first order constraints. Due to this complexity we employ the computer algebra system Mathematica.
It helps with performing the symbolic manipulations necessary to achieve a 
managable final result for $\pi_\mu^{(2)}$. In the course of this subsection we mention the steps for which we used the help of the computer. 

The procedure is analogous to the one in the vacuum case. Some modifications are necessary due to the presence of the electromagnetic field. First, we study the first order diffeomorphism constraints ${}^{(1)}Z^h_{lm}$ and ${}^{(1)}Z^e_{lm}$. The solutions for $\bm y_e$ and $\bm Y_e$ are
\begin{align}
    \bm y_e^{(1)} =& - \frac{1}{\sqrt{l(l+1)}}\Big( -2 \partial_r (\bm y_v) + 2 r \bm y_h - \partial_r \pi_\mu^{(0)} \bm x^v - \frac{1}{2} \pi_\mu^{(0)} \partial_r \bm x^v + \frac{\pi_\lambda^{(0)}}{2r^2} \partial_r \bm x^h\Big)\\
    \bm Y_e^{(1)} =&- \frac{1}{r^2\sqrt{2(l+2)(l-1)}}\qty(-\partial_r(r^2 \bm y_e^{(1)}) +\sqrt{l(l+1)} \qty(\frac{1}{2}\pi_\mu^{(0)} \bm x^v - r^2 \bm y_h +\xi  \alpha))\,,
\end{align}
where we have to substitute $\bm y_e^{(1)}$ in the solution for $\bm Y_e^{(1)}$.
Here and for the rest of the computation we will work in the gauge  $\bm X^e=\bm x^e=0$. 

For the analysis it is convenient to introduce some notation. The following combinations of variables will show up multiple times in the computations of this section:
\begin{align}
    n &:= \frac{1}{2}(l+2)(l-1)\,,\\
    \Delta &:= 1 - \frac{r_s}{r} + \frac{g^2 \xi^2}{4 r^2}\,,\\
    \Lambda &:= n + \frac{3 r_s}{2 r} - \frac{g^2 \xi^2}{2 r^2}\,.
\end{align}

There is one more constraint left. It is the first order Hamiltonian constraint ${}^{(1)} Z^v_{lm}$. The plan is to solve this constraint for $\bm x^h$. Similar to the vacuum case we use a canonical transformation of the gravitational sector to simplify $Z^v$. The transformation removes the derivatives of $\bm x^h$ from the constraints. In addition the transformation helps with putting the final solution for $\pi_\mu^{(2)}$ into 
a more compact form. 

The transformation is very similar to the one in the pure gravity case. The difference is that now $\pi_\mu$ also depends on the charge of the black hole $\xi$. We define new variables $q_1,p_1, q_2, p_2$ by
\begin{align}
    \bm x^v &= q_1 + B q_2 + C \partial_r q_2 + D p_1\\
    \bm x^h &= q_2\\
    \bm y_v &= p_1 + G \partial_r q_2 \\
    \bm y_h &= p_2 - B p_1 + \partial_r \qty[(C - D G) p_1] - \partial_r (G q_1) + K q_2 - BG \partial_r q_2\,,
\end{align}
where we introduced the functions $C,D,G,B$ and $K$. They are defined as 
\begin{align}
    C &:= \frac{1}{r}\\
    D &:= \frac{\pi_\mu^{(0)}}{4 r^2 (\Delta -2)}\\
    G &:= - \frac{\pi_\mu^{(0)}}{4r}\\
    B&:= - \frac{1}{2 r^2 (\Delta -2)}\qty(\frac{r_s}{r} -  (l(l+1)+2))\,.
\end{align}
The function $K$ is longer and given by 
\begin{align}
    K &= \frac{2}{(\Delta - 2)^2  r^2
   \pi_{\mu }^{(0)} } \Bigg(\frac{2}{r} \partial_r \qty(r^2 \Delta  \qty(\Lambda + 2
   \Delta)) -\Lambda  l  (l+1) \qty(\Delta^2 -3 \Delta  + 2) - 2 l(l+1) \qty(2
   \Delta^2 -5 \Delta +4) -4 \Delta \nonumber \\
   &- \frac{r_s}{r} \qty(\Delta^2 -4 \Delta +2)\Bigg)\,.
\end{align} 
In the electromagnetic sector it is convenient to remove the integrals of the momentum $\bm Y^M_e$. To achieve this we introduce new variables $A,\Pi_A$ with a canonical transformation. We define
\begin{align}
    A &:= - \int \bm Y^M_e \dd{r}\\
    \Pi_A &:=- \partial_r \bm X_M^{e}\,.
\end{align}

We insert both transformations into the first order Hamiltonian constraint $Z^v$. We simplify the expression using Mathematica and solve for the variable $q_2$. We obtain
\begin{align}
    q_2^{(1)} &=\frac{1}{r^2 l (l+1) \Lambda}\Big[2 r^4 (\Lambda + 2 \Delta)q_1 - 2 r^5 \qty((\Delta - 2) q_1)' + \sqrt{l(l+1)} g^2 \xi A \Big]\,.
\end{align}

This completes the solution of the first order constraints. The last step of the computation is to insert the expressions for $q_2^{(1)}$, $\bm y_e^{(1)}$ and $\bm Y_e^{(1)}$ into the solution for $\pi_\mu^{(2)}$ in equation \eqref{eq:SolIntegralpimu2}. The result will be a function of the true degrees of freedom $(q_1,p_1)$ and $(A, \Pi_A)$. Using Mathematica for the computation we obtain a result which is still not in a tractable
form. It would be desirable to also write it in terms of potentials similarly to the odd parity sector. 

To achieve this goal we apply two additional canonical transformations. The first transformation is scaling the variable $q_1$ and shifting the variable $p_1$ by contributions proportional to degrees of freedom of the electromagnetic field ($A,\Pi_A$). The goal of this shift of $p_1$ is to remove the coupling between the momentum $p_1$ and the electromagnetic field degrees of freedom $(A,\Pi_A)$. The transformation is given by
\begin{align}
    p_1 &= \sqrt{\frac{(l+2)(l-1)}{l(l+1)}}\frac{r(\Delta - 2)}{\Lambda}\qty(P + \frac{\xi}{r\sqrt{(l+2)(l-1)}}\Pi_A + \Gamma A)\\
    q_1&=\sqrt{\frac{l(l+1)}{(l+2)(l-1)}}\frac{\Lambda}{r (\Delta - 2)}Q \\
    A &= \tilde A - \frac{\xi}{r \sqrt{(l+2)(l-1)}} Q\\
    \Pi_A &= \tilde \Pi_A + \Gamma Q
\end{align}
where we define the function $\Gamma$ as 
\begin{align}
\begin{split}
    \Gamma = \frac{g^2 \xi}{8 \sqrt{(l+2)(l-1)} r \Lambda  (\Delta - 2) \pi_\mu^{(0)}}  \Big(&r^{-8} \pdv{r}\left(16 r^9 \Lambda (1-\Delta)\right) - 8 l(l+1) \left(2 \Delta^2 -11 \Delta + 9\right)\\
    &+16 \Lambda  l(l+1) (1-\Delta) + 16 \left(-4\Delta^2 + \Delta +3  \right)\Big)\,.
\end{split}
\end{align}
With the help of Mathematica we verify that this transformation successfully scales the solution such that $P^2$ and $(Q')^2$ appear both with a factor of $\flatfrac{1}{2}$ in the solution for $\pi_\mu^{(2)}$. As mentioned before all the coupling between the electromagnetic and the gravitational field are removed except for a term of the form $Q A$. 

In the solution we still have couplings between position and momenta of the form $q_1 p_1$ and $\tilde A \tilde \Pi_A$. The next transformation is removing these terms using shifts of the momentum variables $\tilde \Pi_A, \tilde P$. We define the new quantities $Q^e, P^e$ and $A^e, \Pi_A^e$ as
\begin{align}
    Q &= Q^e \\
    P &= P^e + A_{\mathrm{grav}} Q^e\\
    \tilde \Pi_A &= g^2 \Pi_A^e + A_{\mathrm{em}} \frac{1}{g^2}A^e\\
    \tilde A &= \frac{1}{g^2} A^e
\end{align}
In the transformation we introduced two functions, $A_{\mathrm{em}}$ and $A_{\mathrm{grav}}$. The first is defined as
\begin{align}
    A_{\mathrm{em}} = -\frac{g^4 \xi^2 \pi _{\mu }^{(0)}}{8 r^4 \Lambda}\,,
\end{align}
The second one is more complicated and given by
\scriptsize
\begin{align}
    A_{\mathrm{grav}} =& \frac{1}{2 (l-1) (l + 2) r^4 \Lambda (\pi_\mu^{(0)})^{3}(\Delta-2)^2}\Big[64 (l+2)^2(l-1)^2 (3 + l (l + 1) (12 + l(l+1))) r^8 r_s^2 + g^8 \xi^8 (-218 + 85 l (l + 1)) r r_s\nonumber\\
    &- 5 g^{10} \xi^{10} (l+2)(l-1) + 32 (l-1) (l+2) r^7 r_s (4 (1 + l(l+1)) (18 + 5 l(l+1)) r_s^2 - g^2 \xi^2 (l-1) (l + 2) (3 + l (l + 1) (12 + l(l+1))))\nonumber\\
    &+ 2 g^4 \xi^4 r^3 r_s (4 (-934 + 235 l (l + 1)) r_s^2 - g^2 \xi^2 (252 + 5 l (l + 1) (24 + 17 l (l + 1)))) + 2 g^6 \xi^6 r^2 ((914 - 285 l (l + 1)) r_s^2\nonumber\\
    &+ 2 g^2 \xi^2 (8 + 3 l (l + 1) (2 + l(l+1)))) + 8 r^5 r_s (48 (-31 + 5 l(l + 1)) r_s^4 - 8 g^2 \xi^2 (113 + l (l + 1) (7 + 32 l (l + 1))) r_s^2\\
    &+ g^4 \xi^4 (l+2) (l-1) (82 + 17 l (l + 1) (6 + l(l+1)))) + 4 r^6 (48 (35 + l (l + 1) (-1 + 3 l) (4 + 3 l)) r_s^4\nonumber\\
    &- 8 g^2 \xi^2 (l-1) (l + 2) (1 + l(l+1)) (69 + 16 l (l + 1)) r_s^2 + g^4 \xi^4 (l+2)^2(l-1)^2 (4 + l (l + 1) (12 + l(l+1))))\nonumber\\
    &+  4 g^2 \xi^2 r^4 (-8 (-469 + 95 l(l + 1)) r_s^4 + g^2 \xi^2 (724 + l (l + 1) (184 + 223 l (l + 1))) r_s^2\nonumber\\
    &- g^4 \xi^4 (l+2) (l - 1) (16 + l (l + 1) (20 + 3 l (l + 1))))\Big]\,.\nonumber
\end{align}
\normalsize

Mathematica provides us with 
the solution for $\pi_\mu^{(2)}$. We start with the integral in equation \eqref{eq:SolIntegralpimu2}. Then, we insert the solution of the first order constraints and insert all the canonical transformations explained above. The last simplifying step is an integration by parts. The boundary term is recorded in appendix \ref{sec:EvenParityBoundaryTerm}. For now we ignore this term and present the final expression. The Hamiltonian depends on three potentials; one for the gravitational field $V^e_{\mathrm{grav}}$, one for the electromagnetic field  $V^e_{\mathrm{em}}$ and one for the coupling term $V^e_{\mathrm{coup}}$. 

The potentials are defined analogously to the ones in the odd parity sector:
\begin{align}
    V_{\mathrm{coup}}^{e} &:=\frac{g \xi}{r^3} W^e\\
    V_{\mathrm{grav}}^{e} &:= \frac{1}{r^2}\qty(U^e - \frac{3r_s}{2r} W^e)\\
    V_{\mathrm{em}}^{e} &:= \frac{1}{r^2}\qty(U^e + \frac{3 r_s}{2 r} W^e)\,.
\end{align}
The difference is in the definition of the functions $W^e$ and $U^e$. In the even parity case they are more complicated and defined as
\begin{align}
    W^{e} &:= \frac{\Delta}{\Lambda^2} \qty(2 n + \frac{3 r_s}{2 r}) + \frac{1}{\Lambda}\qty(n + \frac{r_s}{2 r})\\
    U^{e} &:= \qty(2 n  + \frac{3 r_s}{2r}) W^e + \qty(\Lambda - n - \frac{r_s}{2 r}) - \frac{2 n \Delta}{\Lambda}
\end{align}
In terms of these definitions the solution for $\pi_\mu^{(2)}$ reads
\begin{align}
    \frac{r}{\pi \pi_\mu^{(0)}} \pi_\mu^{(2)}\Big|_{l\geq2,\text{even}} &= \int \dd{r} \Big[\frac{1}{4 r} \pi_\mu (P^e (Q^e)' + \Pi A') + \frac{1}{2} \qty((P^e)^2 + ({Q^e}')^2 + V_\mathrm{grav}^{e} (Q^e)^2)\\
    &+ \frac{1}{2}\qty(g^2\Pi^2 + \frac{1}{g^2} (A')^2 + \frac{1}{g^2} V_{\mathrm{em}}^{e} A^2) + \frac{\sqrt{(l+2)(l-1)}}{g} V_\mathrm{coup}^{e} A Q^e\Big]
\end{align}

We now consider the boundary term in appendix \ref{sec:EvenParityBoundaryTerm}. For the analysis we use the asymptotic expansion found in equations \eqref{eq:Decay} and \eqref{eq:FAllOffEM}. They imply the following asymptotic behaviour of the variables defined in this section:
\begin{align}
   q_1 &\sim q_1^0 r^{-1} & p_1 &\sim p_1^0 & q_2 &\sim q_2^0 r & A &\sim A_0 r & A^e&\sim A^e_0 r & Q^e &\sim Q^e_0 \,.
\end{align}
The quantities with sub-/superscript $0$ are radial constants but are allowed to vary with respect to $l,m$. Using these definitions the boundary term of $\pi_\mu^{(2)}$ behaves as
\begin{align}
    \begin{split}
    &\frac{1}{\pi \pi_\mu^{(0)}}\Big(- \frac{1}{2}(p_1^0)^2 + \frac{3}{2} (q_1^0)^2 + \frac{(l^2 + l + 2)}{(l+2)(l+1)l(l-1)} (q_1^0)^2 + 2 q_1^0 q_2^0 - \frac{3(l^2+l+2)}{2} q_1^0 q_2^0 + \frac{1}{2} (q_2^0)^2\\
    &- (l^2 + l + 2) (q_2^0)^2+ \frac{1}{8}(3 l^4+ 6 l^3+13 l^2+10l + 16) (q_2^0)^2+ \frac{g^2(l^2+l+2) \xi}{\sqrt{l(l+1)}(l+2)(l-1)} A_0 q_1^0\\
    &- \frac{1}{2} g^2 \xi \sqrt{l(l+1)} q_2^0 A_0 - \frac{1}{2} (Q^e_0)^2 + \frac{g^4 \xi^2}{2(l+2)(l-1)} (A^e_0)^2 + 2 \frac{g^2 \xi}{2 \sqrt{(l+2)(l-1)}}A^e_0 Q^e_0\Big) + O(r^{-1})
    \end{split}
\end{align}
We observe that the leading contributions vanish like $r^{-1/2}$ in the limit $r\to \infty$. This shows that the boundary term is not relevant and we are allowed to drop it in the calculation.

In the rest of this series of papers we are interested in working in the Gullstrand-Painlevé gauge. In this gauge we require $\bm x^v=\bm x^h = \bm x^e = 0$ and have $\bm X^e, \bm Y_e$ and $\bm X^e_M, \bm Y^M_e$ as the true degrees of freedom. In the following we show that it is possible to work in this gauge as well. 
As it turns out we have the same solution for $\pi_\mu^{(2)}$ but now $Q^e,P^e$ will be functions of $\bm X^e$ and $\bm Y^e$ instead. 

As shown in paper \cite{II} we have to first establish weak gauge invariance of the physical Hamiltonian. This is equivalent to showing that the gauge variant contributions to $\pi_\mu^{(2)}$ are a boundary term which vanishes in the limit $r$ to infinity. In this paper we will take a slight variation of this approach which is based on the fact that in both gauges we have $\bm x^e=0$. Since in the end we set $\bm x^e=0$ anyways we only have to worry about the contributions due to $\bm X^e$ and $p_2$. 

The solution of the first order constraints leaving $\bm X^e$ and $p_2$ unfixed is
\begin{align}
\begin{split}
    q_2^{(1)} &= \frac{1}{2 l(l+1) \Lambda}\Big(\sqrt{2(l+2)(l+1)l(l-1)} \bm X^e + 2\qty((l^2 +l + 2) r^2 - 3 r r_s + g^2 \xi^2)q_1\\
    &+ r(4 r^2 + 4 r rs_ - g^2 \xi^2)q_1' + 2 \sqrt{l(l+1)} g^2 \xi A + r^2 \pi_\mu^{(0)} p_2 \Big)
    \end{split}\\
    \begin{split}
    \bm y_e^{(1)} &= \frac{\sqrt{l (l+1)} \pi_\mu^{(0)} \left(\left(l^2+l+6\right) r^2 + 3 r r_s-g^2 \xi ^2\right)}{2 r^3 \left(4 r^2+4 r r_s -g^2 \xi ^2\right)} q_2^{(1)} +\frac{4 \sqrt{l (l+1)} r}{4 r^2+4 r r_s-g^2 \xi ^2} p_1+\frac{2\left(g^2 \xi ^2-6 r r_s\right)}{\sqrt{l(l+1)}r \pi_\mu^{(0)}} q_1\\
    &-\frac{2 r}{\sqrt{l (l+1)}} p_2
    \end{split}\\
\begin{split}
    \bm Y_e^{(1)} &=\frac{\left(2 r_s \left(-g^2 \xi ^2+2 \left(l^2+l+5\right) r^2+4 r r_s\right)-g^2
   \left(l^2+l+2\right) \xi ^2 r\right)}{  r^3 \pi_\mu^{(0)} \left(g^2 \xi ^2-4 r^2-4 r r_s\right)} \bm X^e\\
   &-\frac{8}{2 \sqrt{2(l+2)(l+1)l(l-1)} r^2 (\pi_\mu^{(0)})^3\left(g^2 \xi ^2-4 r^2-4 r r_s\right)}\Big[g^6 \left(l^2+l-6\right) \xi ^6\\
   &~~~~~~~~~~-2 g^4 l (l+1) \left(l^2+l+10\right) \xi ^4 r^2+ 4 r r_s(22-3 l (l+1))  g^4 \xi^4\\
   &~~~~~~~~~~+ 16 \left(l (l+1) \left(l^2+l+10\right)-2\right) r^3 r_s g^2 \xi^2\\
   &~~~~~~~~~~+16 r^2 r_s^2 \left(g^2 (3 l (l+1)-25) \xi ^2-2 \left(l (l+1)
   \left(l^2+l+10\right)-3\right) r^2-4 \left(l^2+l-9\right) r r_s\right)\Big]q_1\\
   &+\frac{2\sqrt{2l(l+1)}\Lambda}{\sqrt{(l+2)(l-1)} \left(-g^2 \xi ^2+4 r^2+4 r r_s\right)}p_1 + \frac{8 \left(l^2+l+3\right)  r r_s-2\left(l^2+l+2\right) g^2 \xi^2}{\sqrt{2(l+2)(l-1)} r^2 \left(g^2 \xi ^2-4 r^2-4 r r_s\right) \pi_\mu^{(0)}} g^2 \xi A\\
   &-\frac{\xi }{\sqrt{2(l+2)(l-1)} r^2}\Pi_A + \frac{4 \left(2 g^2 \xi ^2+\left(l^2+l-6\right) r^2-9 r r_s\right)}{\sqrt{2 (l-1)l (l+1)(l+2)} \left(-g^2 \xi ^2+4 r^2+4 r r_s\right)} p_2
\end{split}
\end{align}
The solutions are then used to define the observable map $O$ which projects the variables $A, \Pi_A$ and $q_1,p_1$ onto gauge invariant functions. The first order solutions $q_2^{(1)},\bm y_e^{(1)},\bm Y_e^{(1)}$ are by definition of first order in the perturbations. Hence, the infinite series in the observable map terminates after the linear order. We define for any function $F$: 
\begin{equation}
    O_F = F + \int \dd{\tilde r}\qty[ p_2(\tilde r)\{q_2(\tilde r) - q_2^{(1)}(\tilde r),F\} + \bm x^e(\tilde r)\{\bm y_e(\tilde r) - \bm y_e^{(1)}(\tilde r),F(r)\} + \bm X^e(\tilde r)\{\bm Y_e(\tilde r) - \bm Y_e^{(1)}(\tilde r),F\}]\,.
\end{equation}
The gauge invariant extensions of $A,\Pi_A$ and $q_1, p_1$ are 
\begin{align}
    O_{A} &= A - \frac{\xi}{\sqrt{2(l+2)(l-1)}r^2} \bm X^e\\
    O_{\Pi_A} &= \Pi_A + \frac{8 \left(l^2+l+3\right)  r r_s-2\left(l^2+l+2\right) g^2 \xi^2}{\sqrt{2(l+2)(l-1)} r^2 \left(g^2 \xi ^2-4 r^2-4 r r_s\right) \pi_\mu^{(0)}} g^2 \xi \bm X^e +\frac{g^2 \xi}{\sqrt{l (l+1)}\Lambda} p_2\\
    O_{q_1} &= q_1 +\sqrt{\frac{2l(l+1)}{(l+2)(l-1)}}\frac{ 2\Lambda}{\left(-g^2 \xi ^2+4 r^2+4 r r_s\right)}\bm X^e\\
    O_{p_1} &=p_1 -\frac{8}{2 \sqrt{2(l+2)(l+1)l(l-1)} r^2 (\pi_\mu^{(0)})^3\left(g^2 \xi ^2-4 r^2-4 r r_s\right)}\Big[g^6 \left(l^2+l-6\right) \xi ^6\nonumber\\
   &~~~~~~~~~~-2 g^4 l (l+1) \left(l^2+l+10\right) \xi ^4 r^2+ 4 r r_s(22-3 l (l+1))  g^4 \xi^4\nonumber\\
   &~~~~~~~~~~+ 16 \left(l (l+1) \left(l^2+l+10\right)-2\right) r^3 r_s g^2 \xi^2\\
   &~~~~~~~~~~+16 r^2 r_s^2 \left(g^2 (3 l (l+1)-25) \xi ^2-2 \left(l (l+1)
   \left(l^2+l+10\right)-3\right) r^2-4 \left(l^2+l-9\right) r r_s\right)\Big]\bm X^e\nonumber\\
   &+\partial_r\qty(\frac{r \left(g^2 \xi ^2-4 r^2-4 r r_s\right)}{2 l (l+1) \Lambda} p_2) - \qty(\frac{2 r^2}{l(l+1)} + \frac{2 r^2}{\Lambda}) p_2\,.\nonumber
\end{align}

We are now fully prepared to investigate the gauge-variant contributions to $\pi_\mu^{(2)}$. In formula \eqref{eq:SolIntegralpimu2} we replace $A,\Pi_A$ as well as $q_1,p_1$ by their gauge invariant extensions $O_A,O_{\pi_A}$ and $O_{q_1},O_{p_1}$. The variables $q_2$, $\bm y_e$, $\bm Y_e$ are replaced by the solutions of the first order constraints. We work in the common gauge $\bm x^e=0$. The result is of the form
\begin{equation}
    \frac{r}{\pi \pi_\mu^{(0)}} \pi_\mu^{(2)} = \int I(O_{q_1},O_{p_1},O_{A},O_{\Pi_A}) \dd{r} + A_1 +A_2 + A_3\,,
\end{equation}
where $I(O_{q_1},O_{p_1},O_{A},O_{\Pi_A})$ is a gauge invariant integrand. $A_1$, $A_2$ and $A_3$ are gauge variant boundary terms and involve $\bm X^e$, $p_2$ and $q_2$. The dependence on $q_2$ of $A_1$ is due to the fact that we simplified the integral for $\pi_\mu^{(2)}$ before inserting the solution for $q_2$. The $A_i$ are shown explicitly in appendix \ref{sec:EvenParityBoundaryTerm}. The asymptotics of the boundary term is determined through the asymptotic behaviour of the canonical variables. The leading contributions based on equations \eqref{eq:Decay} and \eqref{eq:FAllOffEM} are
\begin{align}
    q_1 &\sim q_1^0 r^{-1}, & p_1 &\sim p_1^0, & O_{q_1} &\sim \overline q_1^0 r^{-1}, & O_{p_1} &\sim \overline p_1^0, & q_2 &\sim q_2^0 r, & p_2 &\sim p_2^0 r^{-2}, \nonumber\\
    \bm X^e&\sim \bm X^e_0 r, & A &\sim A_0 r, & O_A&\sim \overline A_0 r\,.
\end{align}
As before the quantities with sub-/superscript $0$ on the right-hand side are independent of $r$. In this notation the leading order contribution to the boundary term is
\begin{align}
    \sum_{i=1}^3 A_i &=\frac{1}{r}\Big[\frac{3 l^4+6 l^3-5 l^2-8 l+8}{16} (\bm X^e_0)^2 -\frac{3}{4} \sqrt{2 (l-1)l(l+1)(l+2)} \bm X^e_0 \overline q^1_0 - \frac{g^2 \xi l(l+1)}{2 \sqrt{2(l+2)(l-1)}} \overline A_0 \nonumber\\
    &+ \frac{2}{(l+2)(l-1)} \overline p_1^0 p_2^0 - \frac{4}{l(l+1)(l-1)^2(l+2)^2} (p_2^0)^2+q_2^0 \Big(\frac{1}{4} \sqrt{2(l-1) l (l+1) (l+2)} \bm X^e_0\\
    &~~~~~~~~~~~~~~~- \frac{1}{8} (3 l(l+1)+2) (l(l+1) q_2^0 - 4 q_1^0) -\frac{1}{2} q_2^0 + \frac{1}{2} \sqrt{l(l+1)} g^2 \xi  A_0\Big)\Big] + O(r^{-3/2})\nonumber
\end{align}
This vanishes as $r^{-1}$ in the limit $r$ to infinity and it is justified to simply express $Q^e,P^e,A^e,\Pi_A^e$ in terms of $\bm X^e, \bm Y^e, A, \Pi_A$. The computation will then yield the same solution for $\pi_\mu^{(2)}$ as before.

We start with the electromagnetic variables. In the analysis of this chapter we defined $A^e,\Pi_A^e$ in terms of the gauge invariants $O_A,O_{\Pi_A}$ using a canonical transformation. To relate to the new gauge we use the explicit formula of the observable map and obtain
\begin{align}
    O_{A} &= A - \frac{\xi}{\sqrt{2(l+2)(l-1)}r^2} \bm X^e\\
    O_{\Pi_A} &= \Pi_A + \frac{8 \left(l^2+l+3\right)  r r_s-2\left(l^2+l+2\right) g^2 \xi^2}{\sqrt{2(l+2)(l-1)} r^2 \left(g^2 \xi ^2-4 r^2-4 r r_s\right) \pi_\mu^{(0)}} g^2 \xi \bm X^e +\frac{g^2 \xi}{\sqrt{l (l+1)}\Lambda} p_2\,,  
\end{align}
where 
\begin{align}
    \label{eq:p2Explicit}
    p_2 &= \bm y_h + B \bm y_v - \partial_r (C \bm y_v)\,.
\end{align}
If we express $\bm y_h, \bm y_v$ in terms of $\bm X^e,\bm Y_e, A,\Pi_A$ we completed the relation of the electromagnetic variables.

For the gravitational perturbations we need to relate $Q^e, P^e$ to $\bm X^e, \bm Y_e,A,\Pi_A$. We already have the relation of $Q^e,P^e$ to $O_{q_1}$ and $O_{p_1}$ from the canonical transformation of the main analysis of this chapter. Using the observable map we found
\begin{equation}
    O_{q_1} = D \bm y_v + \sqrt{\frac{2l(l+1)}{(l+2)(l-1)}} \frac{2 \Lambda}{4 r^2 + 4 r r_s - g^2 \xi^2} \bm X^e
\end{equation}
\begin{align}
    O_{p_1} &=p_1 -\frac{8}{2 \sqrt{2(l+2)(l+1)l(l-1)} r^2 (\pi_\mu^{(0)})^3\left(g^2 \xi ^2-4 r^2-4 r r_s\right)}\Big[g^6 \left(l^2+l-6\right) \xi ^6\nonumber\\
   &~~~~~~~~~~-2 g^4 l (l+1) \left(l^2+l+10\right) \xi ^4 r^2+ 4 r r_s(22-3 l (l+1))  g^4 \xi^4\nonumber\\
   &~~~~~~~~~~+ 16 \left(l (l+1) \left(l^2+l+10\right)-2\right) r^3 r_s g^2 \xi^2\\
   &~~~~~~~~~~+16 r^2 r_s^2 \left(g^2 (3 l (l+1)-25) \xi ^2-2 \left(l (l+1)
   \left(l^2+l+10\right)-3\right) r^2-4 \left(l^2+l-9\right) r r_s\right)\Big]\bm X^e\nonumber\\
   &+\partial_r\qty(\frac{r \left(g^2 \xi ^2-4 r^2-4 r r_s\right)}{2 l (l+1) \Lambda} p_2) - \qty(\frac{2 r^2}{l(l+1)} + \frac{2 r^2}{\Lambda}) p_2\,.\nonumber
\end{align}
where $p_2$ is given by equation \eqref{eq:p2Explicit}.

The last step is to determine the formulas for expressing $\bm y_v$ and $\bm y_h$ in terms of $\bm X^e,\bm Y_e, A, \Pi_A$. Let us start with the solution of $Z^v=0$ for $\bm y_h$ we have
\begin{align}
    \bm y_h= \frac{1}{r^2}\qty(1 - \frac{\pi_\lambda}{\pi_\mu^{(0)}}) \bm y_v - \frac{\sqrt{2(l-1)l(l+1)(l+2)}}{r^2 \pi_\mu^{(0)}} \bm X^e - \frac{2g^2\xi}{r^2\pi_\mu^{(0)}} \sqrt{l(l+1)} A
\end{align}
The solution of $Z^h=0$ for $\bm y_e$ is
\begin{align}
    \bm y_e = \frac{1}{\sqrt{l(l+1)}}\qty(- 2 r \bm y_h + 2 \partial_r \bm y_v)
\end{align}
In this equation we have to replace $\bm y_h$ by its solution. The last constraint $Z^e$ gives a differential equation for $\bm y_v$:
\begin{align}
    \partial_r (r^2 \bm y_e) - \sqrt{l(l+1)} r^2 \bm y_h - \xi \partial_r \bm X^e_M + \sqrt{2(l+2)(l-1)}\qty(r^2 \bm Y_e + \frac{\pi_\lambda^{(0)}}{4 r^2} \bm X^e) =0\,,
\end{align}
where we have to replace $\bm y_e$ and $\bm y_h$ by the above equations. Performing the computation and inserting the explicit formulas for $\pi_\mu^{(0)}$ and $\pi_\lambda^{(0)}$ we obtain
\begin{align}
    2 r^2 \partial_r^2 \bm y_v + \frac{2 r(6 r r_s- g^2 \xi^2)}{4 r r_s - g^2 \xi^2} + \frac{8 (l+2)(l-1) r^2 r_s^2 - 2 (-4 + 3 l(l+1)) r r_s g^2 \xi^2 + (l+2)(l-1) g^4 \xi^4 }{(4 r r_s- g^2 \xi^2)^2} = s(r)\,,
\end{align}
with a ``source'' term $s(r)$ depending on the variables $A, \Pi_A$ and $\bm X^e, \bm Y_e$.
\begin{align}
    s(r) &= - \frac{2\sqrt{2(l-1)l(l+1)(l+2)}r}{\pi_\mu^{(0)}}\bm X^e{}' + \sqrt{2(l-1)l(l+1)(l+2)} r^2 \bm Y_e\nonumber\\
    &+\frac{4\sqrt{2(l-1)l(l+1)(l+2)} \left(-\left(l^2+l-2\right) g^2 \xi^2 r-2 g^2 \xi^2 r_s+4 \left(l^2+l-1\right) r^2 r_s+ 8  r r_s^2\right)}{r (\pi_\mu^{(0)})^3} \bm X^e\\
    &+ \sqrt{l(l+1)} \xi \Pi_A - \frac{4 \sqrt{l(l+1)} r g^2 \xi} {\pi_\mu^{(0)}} A' -\frac{8\sqrt{l(l+1)} g^2 \xi  \left((l+2)(l-1) g^2 \xi ^2-4 \left(l^2+l-1\right) r r_s\right)}{(\pi_\mu^{(0)})^{3}} A\nonumber
\end{align}
The differential equation is linear and of second order with with an 
inhomogeneity given by $s(r)$. The solution of this differential equation 
consists of the sum of a general solution of the homogeneous equation 
(which is a linear function of $r$)
and a particular solution of the inhomogeneous equation.
A particular solution to this differential equation can be found 
simply by quadrature. Without writing the solution explicitly, 
we have succeeded in expressing the quantities 
$Q^e, P^e$ and $A^e, \Pi_A^e$ in terms of $\bm X^e,\bm Y_e$ and $A,\Pi_A$. This shows that the result will be the same in both gauges provided we perform the appropriate canonical transformations.

\subsection{Decoupling of the equations}

For the even and odd parity sector we reduced the physical Hamiltonian to two coupled scalar field Hamiltonians. The potentials are given by $V_{\mathrm{grav}}^{(e/o)}$, $V_{\mathrm{em}}^{(e/o)}$ and $V_{\mathrm{coup}}^{(e/o)}$. In the following we decouple the oscillators through a ``rotation'' of the canonical variables. 

The analysis is completely analogous for both sectors and we drop the labels $e,o$ in this section. We propose the following canonical transformation
\begin{align}
    \mqty(Q^{o/e} \\ A^{e/o}) &= \mqty( \cos \theta & \frac{1}{g} \sin \theta\\ - g \sin \theta & \cos \theta) \mqty( Q_1^{e/o} \\ Q_2^{e/o}) & \mqty(P^{e/o} \\ \Pi_A^{e/o}) &= \mqty(\cos \theta & g \sin \theta \\ - \frac{1}{g} \sin \theta & \cos \theta) \mqty(P_1^{e/o} \\ P_2^{e/o})
\end{align}
The parameter $\theta$ of the transformation is assumed to be a constant depending only on the constants $l$, $r_s$, $g$ and $\xi$. This transformation gives for the solution of $\pi_\mu^{(2)}$ for one of the sectors even or odd:
\begin{align}
\begin{split}
    \frac{r}{\pi \pi_\mu^{(0)}} \pi_\mu^{(2)}\Big|_{l\geq2}=\int \dd{r} &\frac{1}{4r} \pi_{\mu}^{(0)}(P_1 Q_1' + P_2 Q_2')\\
    &+ \frac{1}{2} \Big( P_1^2 + g^2 P_2^2  + (Q_1')^2 + \frac{1}{g^2}(Q_2')^2 + V_1 Q_1^2 + \frac{1}{g^2}V_2 Q_2^2 + \frac{2}{g} \sqrt{(l+2)(l-1)} V_{12} Q_1 Q_2  \Big)\,.
    \end{split}
\end{align}
with the potentials
\begin{align}
    V_1 &:= \frac{1}{r^2}\qty(U - \qty(\frac{3 r_s}{2r} \cos(2 \theta) + \sqrt{(l+2)(l-1)}\frac{g \xi}{r}\sin(2\theta))W)\\
    V_2 &:= \frac{1}{r^2}\qty(U + \qty(\frac{3 r_s}{2 r} \cos(2 \theta) + \sqrt{(l+2)(l-1)} \frac{g \xi}{r} \sin(2\theta))W)\\
    V_{12} &:= \frac{1}{r^2} \qty(\sqrt{(l+2)(l-1)} \frac{g \xi}{r} \cos(2 \theta) - \frac{3 r_s}{2 r} \sin(2 \theta))W
\end{align}
The coupling term $V_{12}$ vanishes provided that
\begin{align}
    \cos(2\theta)^2 &= \frac{9 r_s^2}{9 r_s^2 + 4 (l+2)(l-1) g^2 \xi^2}\,,\\
    \sin(2\theta)^2 &= \frac{4 (l+2)(l-1) g^2 \xi^2}{9 r_s^2 + 4 (l+2)(l-1) g^2 \xi^2}\,.
\end{align}
For taking the square root we need to be careful with the signs.  
We require that the transformation reduces to the identity when the electric charge vanishes,
i.e. when $\xi=0$. This can be achieved by using the positive sign for the cosine. 
The coupling term then only vanishes if we also choose the positive sign for the sine. Thus, $\theta$ is implicitly given by the relations 
\begin{align}
    \cos(2\theta) &= \frac{3 r_s}{\sqrt{9 r_s^2 + 4 (l+2)(l-1) g^2 \xi^2}}\,,\\
    \sin(2\theta) &= \frac{2 \sqrt{(l+2)(l-1)} g \xi}{\sqrt{9 r_s^2 + 4 (l+2)(l-1) g^2 \xi^2}}\,.
\end{align}

In summary, we decoupled the scalar fields in the solution for $\pi_\mu^{(2)}$ in the even and odd parity sector into two independent ones. The potentials for the oscillators are specified by the value of $\theta$ implicitly defined in the above equations. The value of $\theta$ simplifies the potentials $V_1$ and $V_2$ to
\begin{align}
    V_1 &= \frac{1}{r^2}\qty(U - \frac{1}{2r} \sqrt{9 r_s^2 + 4 (l+2)(l-1) g^2 \xi^2} W)\,,\\
    V_2 &= \frac{1}{r^2}\qty(U + \frac{1}{2r} \sqrt{9 r_s^2 + 4 (l+2)(l-1) g^2 \xi^2} W)\,.
\end{align}
We now have four sets of decoupled scalar fields with potentials $V_1^{e/o}$ and $V_2^{e/o}$.

\subsection{Reduced Hamiltonian}

The expression for the reduced Hamiltonian is an explicit expression 
in terms of gravitational variables only. It therefore depends only implicitly on the 
matter content of the theory through the solution of the constraints (which depend on the matter
content) for the gravitational variables. Therefore we can immediately use the results 
of \cite{I,II} and just have to use the matter modified solutions of the constraints. 
We find to second order
\begin{align}
    H=\lim_{r\to \infty} \frac{\pi c}{\kappa r} \pi_\mu^2 &= \lim_{r\to \infty} \frac{\pi }{2 \kappa r}\qty((\pi_\mu^{(0)})^2 + 2 \pi_\mu^{(0)} \pi_\mu^{(2)})\\
    &= M + \frac{1}{\kappa} \int_{\mathbb{R}^+} \dd{r} \frac{1}{4r} \pi_\mu^{(0)} {}^{(2)}C_h + {}^{(2)}C_v\,.
\end{align}
Writing out the contributions explicitly we have
\begin{equation}
    H = M + H_{l=1} + \frac{1}{\kappa} \sum_{l\geq2,m,I} \int_{\mathbb{R}^+} \dd{r} N^3 P^I_{lm} \partial_r Q^I_{lm} + \frac{N}{2}\qty((P^I_{lm})^2 + (\partial_r Q^I_{lm})^2 + V_I (Q^I_{lm})^2)\,,
\end{equation}
where $H_{l=1}$ are the contributions due to the dipole perturbations. We restored the labels $l$ and $m$. $I$ stands for the labels even and odd as well as 1 and 2 from the 
previous chapter. The potentials $V_I$ are the decoupled potentials and $N=1,N^3=4\pi_\mu^{(0)}/r$ are 
the non-vanishing background lapse and shift functions. The dipole part of the physical Hamiltonian is
\begin{align}
    H_{l=1} = \frac{1}{\kappa}\sum_{m} \int \dd{r} \Big[&N^3 \bm Y^M_{o,1m} \bm X^M_{o,1m}{}' + N\Big(\frac{g^2}{2} (\bm Y^M_{o,1m})^2 + \frac{1}{2g^2}\qty((\bm X^{o,1m}_M{}')^2 + \frac{2 + g^2 \xi^2 r^{-2}}{r^2} (\bm X^{o,1m}_M)^2)\nonumber\\
    &+\frac{a_m^2 + 2 \xi a_m \bm X^{o,1m}_M}{2 r^4} \Big)\Big]\nonumber\\
    \sum_m\int \dd{r} &\Big[N^3\Pi_A^{e,1m} A^{e,1m}{}' + N\Big(\frac{1}{2} (\bm y^{1m}_e)^2 - \frac{2 \sqrt{2} \xi}{r^2 \pi_\mu}A^{e,1m} \bm y^{1m}_v + \frac{g^2}{2}(\Pi^{e,1m}_A)^2\nonumber\\
    &+ \frac{1}{2g^2r^2}\qty(2 (A^{e,1m})^2 + r^2 (A^{e,1m}{}')^2)\Big)\Big]
\end{align}
The first integral is from the odd parity and the second integral from the even parity contributions. We start with the interpretation of the odd parity sector and set the integration constant $a_m=0$. We obtain the following integral
\begin{equation}
    \frac{1}{\kappa}\sum_{m} \int \dd{r} \Big[N^3 \bm Y^M_{o,1m} \bm X^M_{o,1m}{}' + N\Big(\frac{g^2}{2} (\bm Y^M_{o,1m})^2 + \frac{1}{2g^2}\qty((\bm X^{o,1m}_M{}')^2 + \frac{2 + g^2 \xi^2 r^{-2}}{r^2} (\bm X^{o,1m}_M)^2)\Big]\,.
\end{equation}
This is the same shape as the other Hamiltonians with $l\geq2$. The potential is given by
\begin{equation}
    V^o_{l=1} = \frac{2 r^2 + g^2 \xi^2}{r^4}\,,
\end{equation}
which is just the evaluation for $l=1$ of $V^o_\mathrm{em}$. 

For the even parity we restrict to the uncharged black hole case. Then, the particular solution of $\bm y_v$ is absent. If furthermore the homogeneous solution is ignored we can remove $\bm y_v$ and $\bm y_e$ in the integral and obtain
\begin{align}
    \sum_m\int \dd{r} &\Big[N^3\Pi_A^{e,1m} A^{e,1m}{}' + N\Big(\frac{g^2}{2}(\Pi^{e,1m}_A)^2\nonumber+ \frac{1}{2g^2r^2}\qty(2 (A^{e,1m})^2 + r^2 (A^{e,1m}{}')^2)\Big)\Big]\,,
\end{align}
The potential in this part of the Hamiltonian is
\begin{align}
    V^e_{l=1}\Big|_{\xi=0} = \frac{2}{r^2}\,.
\end{align}
It is also obtained by considering $V^e_{lm}$ for the case $\xi=0$ and $l=1$. 

\section{Relation to Lagrangian Formalism:}
\label{sec:Comparison}

For a check of consistency we compare our result in the Hamiltonian framework with the Lagrangian formulation discussed in \cite{Chandrasekhar1983}. The Hamiltonian equations of motion need to match the perturbed Lagrangian equations of motion. The part of the physical Hamiltonian which depends on the perturbations is of the form 
\begin{align}
    \int \dd{r} N^3 P Q' + \frac{N}{2}\qty(P^2 + (Q')^2 + V Q^2)\,.
\end{align}
In the pure gravity case, we proved that  the equations of motion are a wave equation
\begin{equation}
    \square Q = V Q\,,
    \label{eq:WaveQ}
\end{equation}
where $\square = g^{ab}D_a D_b$ is the Laplace operator for the  $(\tau,r)$ part of the metric in 
Gullstrand-Painlevé coordinates. $V$ is the relevant potential of the Hamiltonian formulation.

In \cite{Chandrasekhar1983} Chandrasekhar derives the wave equation for the odd and even parity perturbations. He works in the diagonal Schwarzschild coordinates. In order to match the equations of motion we found, we have to perform a change of coordinates. We are in the fortunate situation that the wave equation in \eqref{eq:WaveQ} is written in a covariant form using the Laplace operator $\square$. A change of the coordinate system is straight forward simply by expressing the Laplace operator in the correct coordinates. In the Schwarzschild coordinates the metric takes the form $g = \mathrm{diag}(- \Delta, \Delta^{-1})$ where $\Delta$ is the function of $r_s$ and $\xi$ defined in the section before. Thus, the wave operator $\square$ in Schwarzschild coordinates is given by
\begin{align}
    \square Q = - \Delta^{-1} \partial_t^2 Q + \partial_r (\Delta \partial_r Q) = \Delta^{-1}\qty(- \partial_t^2 + \partial_{r^*}^2) Q\,,
\end{align}
using the tortoise coordinate defined by $\Delta \partial_r = \partial_{r^*}$. The wave equation becomes
\begin{align}
    \qty(- \partial_t^2 + \partial_{r^*}^2) Q = \Delta V Q\,.
\end{align}

In his analysis Chandrasekhar finds the wave equations
\begin{align}
    \qty(- \partial_t^2 + \partial_{r^*}^2) Z_i^{(\pm)} &= V_i^{(\pm)} Z_i^{(\pm)}\,.
\end{align}
In the equation $Z_i^{(\pm)}$ is the master variable for the odd ($-$) and even ($+$) parity perturbations. $i=1,2$ labels to the two independent master functions. $V_i^{(\pm)}$ are the corresponding potentials.

For the odd parity perturbations Chandrasekhar finds
\begin{align}
    V_i^{-} = \Delta \qty[\frac{l(l+1)}{r^2} - \frac{q_j}{r^3}\qty(1 + \frac{q_i}{(l-1)(l+2) r})]\quad \quad (i,j = 1,2, i \neq j)\,,
\end{align}
with $q_1 + q_2 = 3 r_s$ and $-q_1 q_2 = (l+2)(l-1) g^2 \xi^2$. Plugging the expressions for $q_1$ and $q_2$ we find agreement with our odd parity potential $V^o$. 

The even parity potentials are given by
\begin{align}
    V_1^{+} &= \frac{\Delta}{r^2}\qty[U + \frac{1}{2} (q_1 - q_2) W]\\
    V_2^{+} &= \frac{\Delta}{r^2}\qty[U - \frac{1}{2} (q_1 - q_2) W]\,,
\end{align}
where $q_1$ and $q_2$ are defined as in the odd parity case. Inserting $q_1$ and $q_2$, we match the result 
of our calculations. This shows that we recover the known equations in the Hamiltonian formulation. 

\section{Conclusion}

This paper extended the analysis of \cite{II} to include first of all Maxwell matter because 
electromagnetic radiation is one of the most important astrophysical messengers 
next to neutrinos and gravitational waves.
Our formalism delivers the reduced phase space and reduced Hamiltonian perturbatively.
At second order the Hamiltonian decouples and splits into effectively four decoupled 
free scalar field theories with different potentials (i.e. position dependent mass terms). 
Our results match the results obtained previously in the literature. 

The analysis was performed in the Gullstrand-Painlevé gauge. 
In the companion paper \cite{IV} we show how to extend the analysis to 
generalised gauges 
at second order.
In the future we will extend the formalism to include the other matter species of the standard model,
higher orders in the perturbations and most importantly quantum effects.

\appendix

\section{Boundary terms of the even parity computations}
\label{sec:EvenParityBoundaryTerm}
The boundary for the calculation of the even parity solution of $\pi_\mu^{(2)}$ is given by the following expression:
\scriptsize
\begin{align}
    &\frac{-3 g^2 \xi ^2+4 r^2+12 r r_s}{8 r^3} (q_2')^2 - \frac{\left(l^2+l+2\right) r-r_s}{r^3} q_2' q_2 + \qty(-\frac{g^2 \xi ^2}{2 r^2}+\frac{2 r_s}{r}+2) q_1 q_2' + -\frac{3 \sqrt{4 r r_s-g^2 \xi ^2}}{2 r^2} p_1 q_2'\nonumber \\
    &-\frac{\left(\left(l^2+l+2\right) r-r_s\right) \left(g^2 \xi ^2+12 r^2-4 r r_s\right)}{2
   r^2 \left(-g^2 \xi ^2+4 r^2+4 r r_s\right)} q_1 q_2 -\frac{g^2 \sqrt{l(l+1)} \xi }{2 r^3} q_1 \tilde A\nonumber\\
   &+\frac{2 \left(\left(l^2+l+2\right) r-r_s\right) \sqrt{4 r r_s-g^2 \xi ^2} \left(-g^2 \xi
   ^2+12 r^2+4 r r_s\right)}{r^2 \left(-g^2 \xi ^2+4 r^2+4 r r_s\right){}^2} p_1 q_2\nonumber\\
   &+\frac{\sqrt{4 r r_s-g^2 \xi ^2} \left(g^2 \xi ^2-4 r \left(r_s+3 r\right)\right)}{r
   \left(4 r \left(r_s+r\right)-g^2 \xi ^2\right)} p_1 q_1 + \frac{g^2 \xi  \left(g^2 \xi ^2+\left(l^2+l+2\right) r^2-3 r r_s\right)}{\sqrt{l}
   \sqrt{l+1} r \left(-g^2 \xi ^2+\left(l^2+l-2\right) r^2+3 r r_s\right)} A^e q_1\nonumber\\
   &+\qty(\frac{3}{2r}-\frac{32 r^3}{\left(g^2 \xi ^2-4 r \left(r_s+r\right)\right){}^2}) (p_1)^2\nonumber\\
   &-\frac{1}{8 r^5 \left(-g^2 \xi ^2+4 r^2+4 r
   r_s\right){}^2}\Big[-g^6 l (l+1) \xi ^6+g^4 \left(11 l^2+11 l+2\right) \xi ^4 r r_s +8 l (l+1) r^4 \left(2 \left(l^2+l+12\right) r_s^2-g^2 \left(l^2+l+8\right) \xi ^2\right)\nonumber\\
   &~~~~~~~~+8 r^3
   \left(\left(6 l^2+6 l+4\right) r_s^3-g^2 \left(l^4+2 l^3+14 l^2+13 l+2\right) \xi ^2
   r_s\right)+r^2 \left(g^4 l \left(l^3+2 l^2+15 l+14\right) \xi ^4-8 g^2 \left(5 l^2+5
   l+2\right) \xi ^2 r_s^2\right)\nonumber\\
   &~~~~~~~~-16 \left(3 l^4+6 l^3+13 l^2+10 l+16\right) r^6+16
   \left(2 l^4+4 l^3+25 l^2+23 l+18\right) r^5 r_s\Big](q_2)^2\\
   &+\frac{1}{2l (l+1) r \left(-g^2 \xi
   ^2+\left(l^2+l-2\right) r^2+3 r r_s\right){}^2}\Big[-g^6 \xi ^6+2 g^4 \xi ^4 r \left(2 l (l+1) r+5 r_s+r\right)\nonumber\\
   &~~~~~~~~-g^2
   \xi ^2 r^2 \left(2 (l (l+1) (4 l (l+1)-9)-2) r^2+(31 l (l+1)+14) r
   r_s+30 r_s^2\right)\nonumber\\
   &~~~~~~~~+r^3 \left((l-1) (l+2) (l (l+1) (3 l
   (l+1)-4)+4) r^3+(l-1) (l+2) (29 l (l+1)+10) r^2 r_s+3 (19 l
   (l+1)+10) r r_s^2+27 r_s^3\right)\Big] (q_1)^2\nonumber\\
   &+ \frac{g^6 \xi ^4+4 g^4 \xi ^2 r \left(r-r_s\right)}{8 r^3 \left(-g^2 \xi
   ^2+\left(l^2+l-2\right) r^2+3 r r_s\right)} (A^e)^2 + \frac{g^6 \xi ^5-g^4 \xi ^3 r \left(2 \left(l^2+l-4\right) r+3 r_s\right)+4 g^2
   \left(l^2+l-5\right) \xi  r^3 r_s}{\sqrt{l^2+l-2} r^2 \left(4 r \left(r_s+r\right)-g^2
   \xi ^2\right) \left(-g^2 \xi ^2+\left(l^2+l-2\right) r^2+3 r r_s\right)} Q^e A^e\nonumber\\
   &+\frac{1}{8 (l-1)
   (l+2) r^3 \left(4 r r_s-g^2 \xi ^2\right) \left(g^2 \xi ^2-4 r
   \left(r_s+r\right)\right){}^2 \left(-g^2 \xi ^2+\left(l^2+l-2\right) r^2+3 r
   r_s\right)}\Big[g^{10} (6-5 l (l+1)) \xi ^{10}\nonumber\\
   &~~~~~~~~+g^8 (85 l (l+1)-174) \xi ^8 r r_s +2 g^6 \xi ^6 r^2 \left(2 g^2 (l
   (l+1) (3 l (l+1)+7)+6) \xi ^2-3 (95 l (l+1)-278) r_s^2\right)\nonumber\\
   &~~~~~~~~+64 (l-1) (l+2) r^8
   \left(g^2 \left(l^2+l-2\right) \xi ^2+\left(l (l+1) \left(l (l+1)
   \left(l^2+l+10\right)-25\right)-22\right) r_s^2\right)\nonumber\\
   &~~~~~~~~+32 (l-1) (l+2) r^7 r_s \left(4 l
   (l+1) (5 l (l+1)+23) r_s^2-g^2 \left(l (l+1) \left(l (l+1)
   \left(l^2+l+10\right)-23\right)-38\right) \xi ^2\right)\nonumber\\
   &~~~~~~~~+4g^4 (l-1)
   (l+2) \left(l (l+1) \left(l (l+1) \left(l^2+l+10\right)-20\right)-56\right) \xi ^4 r^6\nonumber\\
   &~~~~~~~~-32 g^2 (l (l+1) (l (l+1) (16 l (l+1)+55)-171)-30) \xi^2 r^6 r_s^2
   +192 (l (l+1) (9 l
   (l+1)-8)+43) r_s^4r^6\nonumber\\
   &~~~~~~~~+4 g^2 \xi ^2 r^4 \left(g^4 \left(28-3 l (l+1) \left(l (l+1)
   \left(l^2+l+5\right)-12\right)\right) \xi ^4+g^2 (l (l+1) (223 l (l+1)+156)+668) \xi ^2
   r_s^2-8 (95 l (l+1)-469) r_s^4\right)\nonumber\\
   &~~~~~~~~+8 r^5 r_s \left(g^4 (l (l+1) (l (l+1) (17 l
   (l+1)+72)-198)-84) \xi ^4-8 g^2 (l (l+1) (32 l (l+1)-3)+121) \xi ^2 r_s^2+48 (5 l
   (l+1)-31) r_s^4\right)\nonumber\\
   &~~~~~~~~+2 g^4 \xi ^4 r^3 r_s \left(20 (47 l (l+1)-182) r_s^2-g^2 (l
   (l+1) (85 l (l+1)+128)+204) \xi ^2\right)\nonumber\\
   &~~~~~~~~-256 \left(l^2+l-2\right)^2 r^9 r_s\Big] (Q^e)^2\nonumber
\end{align}
\normalsize

We also analysed the partial gauge invariance of the physical Hamiltonian in the main text. We argued that the gauge variant contributions are equal to a boundary term. With the help of Mathematica we find three different contributions $A_1,A_2,A_3$. The first one has terms proportional to $q_2$ and its derivatives:
\begin{align}
    A_1=&\qty(-\frac{3 g^2 \xi ^2}{4 r^3}+\frac{3 r_s}{r^2}+\frac{1}{r}) (q_2')^2 + \frac{r_s-\left(l^2+l+2\right) r}{r^3} q_2' q_2 + \qty(-\frac{g^2 \xi ^2}{2 r^2}+\frac{2 r_s}{r}+2)q_2' q_1 -\frac{3 \pi_\mu^{(0)}}{4 r^2} q_2' p_1\nonumber\\
    &-\frac{\sqrt{(l-1) l (l+1) (l+2)}}{2 \sqrt{2} r^3} q_2 \bm X^e + -\frac{\left(\left(l^2+l+2\right) r-r_s\right) \left(g^2 \xi ^2+12 r^2-4 r r_s\right)}{8 r^3 \left(r_s+r\right)-2 g^2 \xi ^2 r^2} q_2 q_1 + \frac{\pi_\mu^{(0)}}{4r} q_2 p_2\nonumber\\
   &+\frac{2 \left(\left(l^2+l+2\right) r-r_s\right) \sqrt{4 r r_s-g^2 \xi ^2} \left(4 r \left(r_s+3 r\right)-g^2 \xi
   ^2\right)}{r^2 \left(g^2 \xi ^2-4 r \left(r_s+r\right)\right){}^2} q_2 p_1-\frac{g^2 \sqrt{l (l+1)} \xi }{2 r^3} q_2 A\\
   &+\qty(\frac{64 r^4 \left(r_s-\left(l^2+l+2\right) r\right){}^2}{\left(4 r^2 + 4 r r_s - g^2 \xi ^2\right){}^2}+\frac{l
   (l+1) g^2 \xi ^2-l (l+1) r \left(\left(l^2+l+6\right) r+3 r_s\right)-2 r r_s}{4 r^5}) (q_2)^2\nonumber
\end{align}
The terms involving $\bm X^e$ are captured by $A_2$ and given by
\scriptsize
\begin{align}
    A_2&=- \frac{1}{r^2} (\bm X^e \bm X^e{}') -\frac{g^2 l (l+1) \xi }{2 \sqrt{2(l+2)(l-1)} r^3} \bm X^e O_A + \frac{1}{4 \sqrt{2} \sqrt{(l-1) l (l+1) (l+2)} r^3 \left(4 r r_s-g^2 \xi^2\right) \left(g^2 \xi ^2-4 r \left(r_s+r\right)\right){}^2}\Big[g^8 \left(l^2+l-10\right) \xi ^8\nonumber\\
   &~~~~~~~~~~-2 g^6 \xi ^6 r \left(l (l+1) \left(l^2+l+2\right) r+8 l (l+1) r_s-92 r_s-72 r\right)\nonumber\\
   &~~~~~~~~~~+8 g^4 \xi ^4 r^2 \left(2 \left(l (l+1) \left(l^2+l+19\right)-2\right) r^2+\left(3 l (l+1) \left(l^2+l+2\right)-272\right) r
   r_s+(12 l (l+1)-151) r_s^2\right)\nonumber\\
   &~~~~~~~~~~+32 g^2 \xi ^2 r^3 \left(-4 \left(l (l+1) \left(l^2+l+19\right)-9\right) r^2 r_s-\left(3
   l (l+1) \left(l^2+l+2\right)-314\right) r r_s^2+3 (l-1) l (l+1) (l+2) r^3-2 (4 l (l+1)-53) r_s^3\right)\nonumber\\
   &~~~~~~~~~~+128 r^4 r_s
   \left(\left(2 l (l+1) \left(l^2+l+19\right)-27\right) r^2 r_s+\left(l (l+1) \left(l^2+l+2\right)-114\right) r r_s^2-3
   (l-1) l (l+1) (l+2) r^3+(2 l (l+1)-27) r_s^3\right) \Big] \bm X^e O_{q_1}\nonumber\\
    &\frac{1}{2 \sqrt{2} \sqrt{l (l+1)
   \left(l^2+l-2\right)} r^3 \left(4 r r_s-g^2 \xi ^2\right){}^{3/2} \left(4 r \left(r_s+r\right)-g^2 \xi ^2\right){}^3}\Big[-g^{10} \left(l^2+l+18\right) \xi ^{10}+8 g^8 (2 l (l+1)+51) \xi ^8 r r_s\nonumber\\
   &~~~~~~~~~~+256 r^7 r_s \left(g^2 (2-l (l+1) (7 l(l+1)-2)) \xi ^2+4 l (l+1) \left(l^2+l-28\right) r_s^2\right)\nonumber\\
   &~~~~~~~~~~+32 r^3 \left(g^6 \left(l (l+1)
   \left(l^2+l+16\right)-87\right) \xi ^6 r_s+g^4 (8 l (l+1)+486) \xi ^4 r_s^3\right)\nonumber\\
   &~~~~~~~~~~+2 g^6 \xi ^6 r^2 \left(-g^2
   \left(l^2+l-4\right) \left(l^2+l+18\right) \xi ^2-24 (2 l (l+1)+75) r_s^2\right)\nonumber\\
   &~~~~~~~~~~-32 r^6 \left(-g^4 l (l+1) (7 l (l+1)-2)  \xi ^4+8 g^2 \left(3 l (l+1) \left(l^2+l-27\right)-4\right) \xi ^2 r_s^2+16 \left(l (l+1) \left(l^2+l+22\right)-117\right)
   r_s^4\right)\nonumber\nonumber\\
   &~~~~~~~~~~+64 r^5 \left(g^4 \left(3 l (l+1) \left(l^2+l-26\right)-10\right) \xi ^4 r_s+4 g^2 \left(2 l (l+1)
   \left(l^2+l+20\right)-219\right) \xi ^2 r_s^3+432 r_s^5\right)\nonumber\nonumber\\
   &~~~~~~~~~~-16 r^4 \left(g^6 \left(l (l+1)
   \left(l^2+l-25\right)-6\right) \xi ^6+6 g^4 \left(2 l (l+1) \left(l^2+l+18\right)-199\right) \xi ^4 r_s^2 +16 g^2
   \left(l^2+l+129\right) \xi ^2 r_s^4\right)\\
   &~~~~~~~~~~+512 (l (l+1) (7 l (l+1)-2)-3) r^8 r_s^2 \Big]\bm X^e O_{p_1}\nonumber\nonumber\\
   &+\frac{1}{2(l-1) l (l+1) (l+2) r^5 (\pi_\mu^{(0)})^3 \left(g^2 \xi ^2-4 r \left(r+r_s\right)\right){}^4}\Big\{16 r^6 \left(4 r r_s-g^2 \xi ^2\right){}^3 \left(-5 g^4 \xi ^4+8 g^2 r \left(3 r+5 r_s\right) \xi ^2+16 r^2 \left(3
   r^2-6 r_s r-5 r_s^2\right)\right) l^8\nonumber\\
   &~~~~~~~~~~+64 r^6 \left(4 r r_s-g^2 \xi ^2\right){}^3 \left(-5 g^4 \xi ^4+8 g^2 r \left(3 r+5
   r_s\right) \xi ^2+16 r^2 \left(3 r^2-6 r_s r-5 r_s^2\right)\right) l^7\nonumber\\
   &~~~~~~~~~~+4 r^2 \left(4 r r_s-g^2 \xi ^2\right){}^3 \Big[-2
   g^8 \xi ^8+g^6 r \left(2 r+29 r_s\right) \xi ^6-4 g^4 r^2 \left(56 r^2-5 r_s r+39 r_s^2\right) \xi ^4+16 g^2 r^3
   \left(r+r_s\right) \left(146 r^2-39 r_s r+23 r_s^2\right) \xi ^2\nonumber\\
   &~~~~~~~~~~~~~~~~~~~~+64 r^4 \left(4 r^4-149 r_s r^3-51 r_s^2 r^2+9 r_s^3 r-5
   r_s^4\right)\Big] l^6\nonumber\\
   &~~~~~~~~~~-4 r^2 \left(4 r r_s-g^2 \xi ^2\right){}^3 \Big[6 g^8 \xi ^8-3 g^6 r \left(2 r+29 r_s\right) \xi
   ^6+4 g^4 r^2 \left(98 r^2-15 r_s r+117 r_s^2\right) \xi ^4\nonumber\\
   &~~~~~~~~~~~~~~~~~~~~-16 g^2 r^3 \left(354 r^3+181 r_s r^2-48 r_s^2 r+69 r_s^3\right)
   \xi ^2+64 r^4 \left(30 r^4+363 r_s r^3+83 r_s^2 r^2-27 r_s^3 r+15 r_s^4\right)\Big] l^5\nonumber\\
   &~~~~~~~~~~+\left(g^2 \xi ^2-4 r
   r_s\right){}^2 \Big[-g^{12} \xi ^{12}+10 g^{10} r \left(3 r_s-2 r\right) \xi ^{10}+4 g^8 r^2 \left(350 r^2+81 r_s r-90
   r_s^2\right) \xi ^8\nonumber\\
   &~~~~~~~~~~~~~~~~~~~~-16 g^6 r^3 \left(793 r^3+1513 r_s r^2+127 r_s^2 r-140 r_s^3\right) \xi ^6-64 g^4 r^4 \left(64 r^4-2582
   r_s r^3-2432 r_s^2 r^2-95 r_s^3 r+120 r_s^4\right) \xi ^4\nonumber\\
   &~~~~~~~~~~~~~~~~~~~~-256 g^2 r^5 \left(17 r^5-83 r_s r^4+2754 r_s^2 r^3+1725 r_s^3
   r^2+33 r_s^4 r-54 r_s^5\right) \xi ^2\nonumber\\
   &~~~~~~~~~~~~~~~~~~~~+1024 r^6 r_s \left(17 r^5-32 r_s r^4+965 r_s^2 r^3+456 r_s^3 r^2+4 r_s^4 r-10 r_s^5\right)\Big] l^4\nonumber\\
   &~~~~~~~~~~+2 \left(g^2 \xi ^2-4 r r_s\right){}^2 \Big[-g^{12} \xi ^{12}+10 g^{10} r \left(3 r_s-4 r\right)
   \xi ^{10}+2 g^8 r^2 \left(710 r^2+347 r_s r-180 r_s^2\right) \xi ^8\nonumber\\
   &~~~~~~~~~~~~~~~~~~~~-8 g^6 r^3 \left(1726 r^3+3011 r_s r^2+594 r_s^2 r-280
   r_s^3\right) \xi ^6+32 g^4 r^4 \left(434 r^4+5559 r_s r^3+4759 r_s^2 r^2+500 r_s^3 r-240 r_s^4\right) \xi ^4\nonumber\\
   &~~~~~~~~~~~~~~~~~~~~-128 g^2 r^5 \left(98 r^5+973 r_s r^4+5878 r_s^2 r^3+3325 r_s^3 r^2+206 r_s^4 r-108 r_s^5\right) \xi ^2\nonumber\\
   &~~~~~~~~~~~~~~~~~~~~+512 r^6 r_s \left(98 r^5+513
   r_s r^4+2045 r_s^2 r^3+867 r_s^3 r^2+33 r_s^4 r-20 r_s^5\right)\Big] l^3\nonumber\\
   &~~~~~~~~~~-\left(g^2 \xi ^2-4 r r_s\right){}^2 \Big[-7
   g^{12} \xi ^{12}+2 g^{10} r \left(71 r_s-190 r\right) \xi ^{10}+40 g^8 r^2 \left(114 r^2+230 r_s r-25 r_s^2\right) \xi^8\nonumber\\
   &~~~~~~~~~~~~~~~~~~~~+16 g^6 r^3 \left(356 r^3-5468 r_s r^2-5487 r_s^2 r+149 r_s^3\right) \xi ^6+64 g^4 r^4 \left(56 r^4-860 r_s r^3+9501
   r_s^2 r^2+6424 r_s^3 r+53 r_s^4\right) \xi ^4\nonumber\\
   &~~~~~~~~~~~~~~~~~~~~-256 g^2 r^5 \left(24 r^5+74 r_s r^4-679 r_s^2 r^3+7119 r_s^3 r^2+3687 r_s^4
   r+95 r_s^5\right) \xi ^2\nonumber\\
   &~~~~~~~~~~~~~~~~~~~~+1024 r^6 r_s \left(24 r^5+29 r_s r^4-186 r_s^2 r^3+1946 r_s^3 r^2+830 r_s^4 r+29 r_s^5\right)\Big] l^2\nonumber\\
   &~~~~~~~~~~-4 \left(g^2 \xi ^2-4 r r_s\right){}^2 \Big[-2 g^{12} \xi ^{12}+g^{10} r \left(43 r_s-106 r\right)\xi ^{10}+4 g^8 r^2 \left(374 r^2+623 r_s r-85 r_s^2\right) \xi ^8\nonumber\\
   &~~~~~~~~~~~~~~~~~~~~-4 g^6 r^3 \left(520 r^3+6972 r_s r^2+5818 r_s^2 r-289
   r_s^3\right) \xi ^6+16 g^4 r^4 \left(328 r^4+1956 r_s r^3+11870 r_s^2 r^2+6705 r_s^3 r-67 r_s^4\right) \xi ^4\nonumber\\
   &~~~~~~~~~~~~~~~~~~~~-64 g^2 r^5
   \left(80 r^5+672 r_s r^4+2294 r_s^2 r^3+8769 r_s^3 r^2+3804 r_s^4 r+41 r_s^5\right) \xi ^2\nonumber\\
   &~~~~~~~~~~~~~~~~~~~~+256 r^6 r_s \left(80 r^5+342
   r_s r^4+847 r_s^2 r^3+2375 r_s^3 r^2+849 r_s^4 r+19 r_s^5\right)\Big] l\nonumber\\
   &~~~~~~~~~~-4 \left(-3 g^6 \xi ^6+44 g^4 r r_s \xi ^4-8 g^2
   r^2 r_s \left(2 r+25 r_s\right) \xi ^2+48 r^3 r_s^2 \left(r+6 r_s\right)\right){}^2 \left(-3 g^4 \xi ^4+24 g^2 r
   \left(r+r_s\right) \xi ^2+16 r^2 \left(r^2-6 r_s r-3 r_s^2\right)\right)\Big\}(\bm X^e)^2\nonumber\\
   &-\frac{1}{r^3}(\bm X^e)^2\nonumber\\
   &+\frac{1}{2 (l (l+1))^{3/2} \left(4 r r_s-g^2 \xi ^2\right) \left(g^2 \xi ^2-4 r
   \left(r_s+r\right)\right){}^2 \left(-g^2 \xi ^2+\left(l^2+l-2\right) r^2+3 r r_s\right)
   \sqrt{\left(l^2+l-2\right) \left(8 r r_s-2 g^2 \xi ^2\right)}}\Big[-g^{10} \left(l^2+l+18\right) \xi ^{10}\nonumber\\
   &~~~~~~~~~~-2 g^8 \xi ^8 r \left(\left(l^2+l-4\right)
   \left(l^2+l+18\right) r-4 (2 l (l+1)+51) r_s\right)\nonumber\\
   &~~~~~~~~~~-16 g^6 \xi ^6 r^2 \left(\left(l (l+1)
   \left(l^2+l-25\right)-6\right) r^2-2 \left(l (l+1) \left(l^2+l+16\right)-87\right) r r_s+3 (2
   l (l+1)+75) r_s^2\right)\nonumber\\
   &~~~~~~~~~~+32 g^4 \xi ^4 r^3 \left(2 \left(3 l (l+1)
   \left(l^2+l-26\right)-10\right) r^2 r_s-3 \left(2 l (l+1) \left(l^2+l+18\right)-199\right) r
   r_s^2+l (l+1) (7 l (l+1)-2) r^3+2 (4 l (l+1)+243) r_s^3\right)\nonumber\\
   &~~~~~~~~~~-256 g^2 \xi ^2 r^4 r_s
   \left(\left(3 l (l+1) \left(l^2+l-27\right)-4\right) r^2 r_s-\left(2 l (l+1)
   \left(l^2+l+20\right)-219\right) r r_s^2+\left(l^2+l+129\right) r_s^3+(l (l+1) (7 l
   (l+1)-2)-2) r^3\right)\nonumber\\
   &~~~~~~~~~~+512 r^5 r_s^2 \left(2 l (l+1) \left(l^2+l-28\right) r^2 r_s-\left(l
   (l+1) \left(l^2+l+22\right)-117\right) r r_s^2+(l (l+1) (7 l (l+1)-2)-3) r^3+54
   r_s^3\right)\Big] \bm X^e p_2'\nonumber\\
   &+ \frac{\sqrt{(l-1) (l+1) (l+2)}}{2 \sqrt{2} (l+1)^2 \left(l^2+l-2\right) r
   \left(\left(l^2+l-2\right) r^2+3 r_s r-g^2 \xi ^2\right){}^2 \left(l \left(4 r r_s-g^2 \xi
   ^2\right)\right){}^{3/2} \left(4 r \left(r+r_s\right)-g^2 \xi ^2\right){}^3}\Big[-3 g^{14} \left(l^2+l+30\right) \xi ^{14}\nonumber\\
   &~~~~~~~~~~+2 g^{12} (29 l
   (l+1)+1326) r r_s \xi ^{12}-g^{10} r^2 \left(g^2 (l (l+1) (5 l (l+1)+112)-900) \xi ^2+4 (103 l
   (l+1)+8211) r_s^2\right) \xi ^{10}\nonumber\\
   &~~~~~~~~~~+4 g^8 r^3 r_s \left(8 g^2 (l (l+1) (4 l (l+1)+87)-720) \xi
   ^2+9 (29 l (l+1)+6172) r_s^2\right) \xi ^8\nonumber\\
   &~~~~~~~~~~+2 g^6 r^4 \Big(-g^4 (l-3) (l+4) (l (l+1) (5 l
   (l+1)+76)-28) \xi ^4\nonumber\\
   &~~~~~~~~~~~~~~~~~~~~-6 g^2 (l (l+1) (113 l (l+1)+2340)-19768) r_s^2 \xi ^2+16 (58 l
   (l+1)-27759) r_s^4\Big) \xi ^6\nonumber\\
   &~~~~~~~~~~-1024 (l-1) (l+2) (l (l+1) (l (l+1) (5 l (l+1)-29)+11)+12)
   r^{12} r_s^2\nonumber\\
   &~~~~~~~~~~+256 r^{11} r_s \Big(g^2 (l-1) (l+2) (l (l+1) (l (l+1) (10 l (l+1)-57)+22)+16)
   \xi ^2\nonumber\\
   &~~~~~~~~~~~~~~~~~~~~+4 (l (l+1) (l (l+1) (l (l+1) (2 l (l+1)+95)-334)+404)-12) r_s^2\Big)\nonumber\\
   &~~~~~~~~~~+64 r^{10}
   \Big(-g^4 (l-1) l (l+1) (l+2) (l (l+1) (5 l (l+1)-28)+12) \xi ^4\nonumber\\
   &~~~~~~~~~~~~~~~~~~~~-8 g^2 (l (l+1) (l (l+1) (l
   (l+1) (3 l (l+1)+134)-423)+547)-6) r_s^2 \xi ^2\nonumber\\
   &~~~~~~~~~~~~~~~~~~~~+16 \left(l (l+1) \left(l (l+1)
   \left(l^2+l-5\right) (3 l (l+1)+52)+579\right)-1017\right) r_s^4\Big)\nonumber\\
   &~~~~~~~~~~+64 r^9 \Big(16 (l
   (l+1) (l (l+1) (5 l (l+1)-86)-757)+36) r_s^5\nonumber\\
   &~~~~~~~~~~~~~~~~~~~~-4 g^2 (l (l+1) (l (l+1) (l (l+1) (12 l
   (l+1)+145)-704)+1153)-3848) \xi ^2 r_s^3\nonumber\\
   &~~~~~~~~~~~~~~~~~~~~+g^4 (l (l+1) (l (l+1) (l (l+1) (6 l
   (l+1)+251)-688)+956)+80) \xi ^4 r_s\Big)\nonumber\\
   &~~~~~~~~~~+4 r^5 \Big(g^8 (l (l+1) (3 l (l+1) (15 l
   (l+1)+16)-8548)+4000) r_s \xi ^8\nonumber\\
   &~~~~~~~~~~~~~~~~~~~~+4 g^6 (7 l (l+1) (68 l (l+1)+1315)-79032) r_s^3 \xi ^6-32 g^4
   (133 l (l+1)-16413) r_s^5 \xi ^4\Big)\nonumber\\
   &~~~~~~~~~~-32 r^8 \Big(g^6 \left(l (l+1) \left(l (l+1) \left(l
   (l+1) \left(l^2+l+39\right)-88\right)+132\right)+48\right) \xi ^6\nonumber\\
   &~~~~~~~~~~~~~~~~~~~~-2 g^4 (l (l+1) (l (l+1) (3 l
   (l+1) (6 l (l+1)+71)-556)-6)-5408) r_s^2 \xi ^4\nonumber\\
   &~~~~~~~~~~~~~~~~~~~~+16 g^2 (l (l+1) (3 l (l+1) (5 l
   (l+1)-51)-2578)+906) r_s^4 \xi ^2+288 (l (l+1) (3 l (l+1)+41)-375) r_s^6\Big)\nonumber\\
   &~~~~~~~~~~+4 r^6
   \Big(g^8 (l (l+1) (l (l+1) (l (l+1) (3 l (l+1)+34)+72)-568)-816) \xi ^8\nonumber\\
   &~~~~~~~~~~~~~~~~~~~~-8 g^6 (l (l+1) (l(l+1) (40 l (l+1)-67)-7549)+3864) r_s^2 \xi ^6-16 g^4 (l (l+1) (374 l (l+1)+6615)-57687) r_s^4
   \xi ^4\nonumber\\
   &~~~~~~~~~~~~~~~~~~~~+1152 g^2 (8 l (l+1)-591) r_s^6 \xi ^2\Big)\nonumber\\
   &~~~~~~~~~~-16 r^7 \Big(1728 \left(l^2+l-54\right)
   r_s^7-48 g^2 (l (l+1) (52 l (l+1)+821)-7298) \xi ^2 r_s^5\nonumber\\
   &~~~~~~~~~~~~~~~~~~~~+4 g^4 (-l (l+1) (10 l (l+1) (7 l
   (l+1)-37)-12829)-6176) \xi ^4 r_s^3\nonumber\\
   &~~~~~~~~~~~~~~~~~~~~+g^6 (l (l+1) (l (l+1) (l (l+1) (12 l
   (l+1)+139)-40)-1148)-3392) \xi ^6 r_s\Big)\Big]\bm X^e p_2\nonumber
\end{align}
\normalsize
The remaining terms are the contributions due to $p_2$. They are given by $A_3$:
\scriptsize
\begin{align}
    A_3=&\frac{r^5 \left(3 g^4 \xi ^4-24 g^2 \xi ^2 r \left(r_s+r\right)-16 r^2 \left(r^2-6 r r_s-3
   r_s^2\right)\right)}{2 l^2 (l+1)^2 \left(-g^2 \xi ^2+\left(l^2+l-2\right) r^2+3 r
   r_s\right){}^2} (p_2')^2\nonumber\\
   &+\frac{1}{8 l^2 (l+1)^2 r^2 \left(4 r \left(r_s+r\right)-g^2 \xi ^2\right) \Lambda^3} \Big[\left(-3 g^4 \xi ^4+24 g^2 \xi ^2 r \left(r_s+r\right)+16 r^2 \left(r^2-6 r r_s-3
   r_s^2\right)\right)\times\nonumber\\
   &~~~~~~~~~~\times\left(-5 g^4 \xi ^4+g^2 \xi ^2 r \left(\left(l^2+l+10\right) r+34
   r_s\right)+2 r^2 \left((l-1) (l+2) \left(l^2+l-4\right) r^2-4 \left(l^2+l+1\right) r r_s-27
   r_s^2\right)\right)\Big] p_2 p_2'\nonumber\\
   &+ \frac{r^2 \sqrt{4 r r_s-g^2 \xi ^2} \left(g^2 \xi ^2-4 r \left(r_s+3 r\right)\right)}{l (l+1)
   \left(-g^2 \xi ^2+\left(l^2+l-2\right) r^2+3 r r_s\right)} p_2' O_{q_1} + \frac{r^2 \left(3 g^4 \xi ^4-24 g^2 \xi ^2 r \left(r_s+r\right)-16 r^2 \left(r^2-6 r r_s-3
   r_s^2\right)\right)}{l (l+1) \left(4 r \left(r_s+r\right)-g^2 \xi ^2\right) \left(-g^2 \xi
   ^2+\left(l^2+l-2\right) r^2+3 r r_s\right)} p_2' O_{p_1}\nonumber\\
   &+ \frac{1}{4 l (l+1) r^3 \pi_\mu^{(0)} \left(4 r \left(r_s+r\right)-g^2 \xi ^2\right) \Lambda^2}\Big[-9 g^8 \xi ^8+g^6 \xi ^6 r \left((5 l (l+1)+142) r+131 r_s\right)\nonumber\\
   &~~~~~~~~~~-2 g^4 \xi ^4 r^2
   \left(28 \left(l^2+l+4\right) r^2+(31 l (l+1)+750) r r_s+353 r_s^2\right)\nonumber\\
   &~~~~~~~~~~-16 g^2 \xi ^2 r^3
   \left((l-1) (l+2) (2 l (l+1)-13) r^3-(33 l (l+1)+65) r^2 r_s-(16 l (l+1)+321) r r_s^2-104
   r_s^3\right)\nonumber\\
   &~~~~~~~~~~+32 r^4 r_s \left(-11 \left(l^2+l+16\right) r r_s^2+(l-1) (l+2) (4 l (l+1)-27)
   r^3-(38 l (l+1)+17) r^2 r_s-45 r_s^3\right)\Big] O_{q_1} p_2\nonumber\\
   & +\frac{1}{4 l (l+1) r^3 \left(g^2 \xi ^2-4 r \left(r_s+r\right)\right){}^2 \Lambda^2} \Big[ \left(-3 g^4 \xi ^4+24 g^2 \xi ^2 r \left(r_s+r\right)+16 r^2 \left(r^2-6 r r_s-3
   r_s^2\right)\right)\times \\
   &~~~~~~~~~~\times\left(-5 g^4 \xi ^4+g^2 \xi ^2 r \left(\left(l^2+l+10\right) r+34
   r_s\right)+2 r^2 \left((l-1) (l+2) \left(l^2+l-4\right) r^2-4 \left(l^2+l+1\right) r r_s-27
   r_s^2\right)\right)\Big] O_{p_1} p_2\nonumber\\
   &+ \frac{1}{32 l^2 (l+1)^2r^5 \left(g^2 \xi ^2-4 r \left(r+r_s\right)\right){}^2 \Lambda^4}\Big[ 128 \left(l^2+l-2\right)^2 (3 l (l+1)-8) r^{12}\nonumber\\
   &~~~~~~~~~~+64 (l-1) (l+2) (l (l+1) (l (l+1)
   (7 l (l+1)-47)+154)-224) r_s r^{11}\nonumber\\
   &~~~~~~~~~~+16 \Big(4 (2 l (l+1)-7) \left(l (l+1) \left(l (l+1)
   \left(l^2+l-31\right)+98\right)-32\right) r_s^2\nonumber\\
   &~~~~~~~~~~~~~~~~~~~~-g^2 (l-1) (l+2) (l (l+1) (l (l+1) (7 l
   (l+1)-48)+196)-352) \xi ^2\Big) r^{10}\nonumber\\
   &~~~~~~~~~~+16 r_s \Big(g^2 (992-l (l+1) (l (l+1) (l (l+1) (4 l
   (l+1)-111)+540)-460)) \xi ^2\nonumber\\
   &~~~~~~~~~~~~~~~~~~~~-4 \left(l (l+1) \left(l (l+1) \left(l (l+1)
   \left(l^2+l+27\right)+57\right)-1817\right)+2904\right) r_s^2\Big) r^9\nonumber\\
   &~~~~~~~~~~+8 \Big(g^4 \left(l
   \left(l \left((l (l (l+4)-15)-59) l^3+103 l+242\right)+180\right)-744\right) \xi ^4\nonumber\\
   &~~~~~~~~~~~~~~~~~~~~+2 g^2
   \left(l (l+1) \left(3 l (l+1) \left(l (l+1)
   \left(l^2+l+24\right)+191\right)-7778\right)+11032\right) r_s^2 \xi ^2\nonumber\\
   &~~~~~~~~~~~~~~~~~~~~-24 (l (l+1) (l (l+1) (3 l (l+1)+35)-803)+227) r_s^4\Big) r^8\nonumber\\
   &~~~~~~~~~~+4 r_s \Big(-g^4 \left(l (l+1) \left(3 l (l+1) \left(l
   (l+1) \left(l^2+l+21\right)+312\right)-9716\right)+12160\right) \xi ^4\nonumber\\
   &~~~~~~~~~~~~~~~~~~~~+12 g^2 (l (l+1) (l
   (l+1) (13 l (l+1)+186)-3517)-696) r_s^2 \xi ^2-432 \left(l^2+l-31\right) \left(l^2+l+6\right)
   r_s^4\Big) r^7\nonumber\\
   &~~~~~~~~~~+\Big(g^6 \left(l (l+1) \left(l (l+1) \left(l (l+1)
   \left(l^2+l+18\right)+420\right)-3752\right)+4000\right) \xi ^6\nonumber\\
   &~~~~~~~~~~~~~~~~~~~~-4 g^4 (3 l (l+1) (7 l (l+1) (3
   l (l+1)+49)-5534)-13664) r_s^2 \xi ^4\nonumber\\
   &~~~~~~~~~~~~~~~~~~~~+48 g^2 (l (l+1) (51 l (l+1)-1132)-10858) r_s^4 \xi
   ^2-1728 \left(l^2+l-81\right) r_s^6\Big) r^6\nonumber\\
   &~~~~~~~~~~+g^2 \xi ^2 r_s \Big(g^4 (l (l+1) (3 l (l+1)
   (15 l (l+1)+268)-11212)-18176) \xi ^4\nonumber\\
   &~~~~~~~~~~~~~~~~~~~~+12 g^2 (27416-l (l+1) (115 l (l+1)-2213)) r_s^2 \xi
   ^2+432 (7 l (l+1)-570) r_s^4\Big) r^5\nonumber\\
   &~~~~~~~~~~+g^4 \xi ^4 \Big(-g^4 \left(l^2+l-10\right) \left(3 l
   (l+1) \left(l^2+l+29\right)+182\right) \xi ^4\nonumber\\
   &~~~~~~~~~~~~~~~~~~~~+g^2 (l (l+1) (387 l (l+1)-6302)-101256) r_s^2
   \xi ^2-12 (183 l (l+1)-14857) r_s^4\Big) r^4\nonumber\\
   &~~~~~~~~~~+g^6 \xi ^6 r_s \left(2 g^2 (7620-l (l+1) (27 l
   (l+1)-362)) \xi ^2+7 (121 l (l+1)-9720) r_s^2\right) r^3\nonumber\\
   &~~~~~~~~~~+g^8 \xi ^8 \left(g^2 (l (l+1) (3 l
   (l+1)-32)-900) \xi ^2-3 (61 l (l+1)-4816) r_s^2\right) r^2+3 g^{10} (7 l (l+1)-540) \xi ^{10}
   r_s r-g^{12} \left(l^2+l-75\right) \xi ^{12}\Big] (p_2)^2\nonumber
\end{align}


\end{document}